\documentclass[preprint,12pt,authoryear]{elsarticle}

\usepackage{amssymb}
\usepackage{amsmath}

\usepackage{lineno}

\usepackage{graphicx}
\usepackage{subfigure}
\usepackage{array}
\usepackage{url}
\usepackage{color}
\usepackage{hyperref}
\usepackage{multirow}
\usepackage{subfloat}
\usepackage{booktabs} 

\usepackage{algorithm}
\usepackage{algpseudocode}

\usepackage{utfsym}

\journal{Expert Systems with Applications}

\begin{document}

\begin{frontmatter}



\title{BandCondiNet: Parallel Transformers-based Conditional Popular Music Generation with Multi-View Features} 


\author[label1]{Jing~Luo}
\ead{luojingl@stu.xjtu.edu.cn}

\author[label1]{Xinyu~Yang\corref{cor1}}
\ead{yxyphd@mail.xjtu.edu.cn}

\author[label2]{Dorien~Herremans}
\ead{dorien_herremans@sutd.edu.sg}

\cortext[cor1]{Corresponding author: Xinyu~Yang}

\affiliation[label1]{organization={Xi'an Jiaotong University},
            addressline={28 Xianning West Rd}, 
            city={Xi'an},
            postcode={710049}, 
            country={China}}
\affiliation[label2]{organization={Singapore University of Technology and Design},
	addressline={8 Somapah Rd}, 
	postcode={487372}, 
	country={Singapore}}

\begin{abstract}
Conditional music generation offers significant advantages in terms of user convenience and control, presenting great potential in AI-generated content research. However, building conditional generative systems for multitrack popular songs presents three primary challenges: insufficient fidelity of input conditions, poor structural modeling, and inadequate inter-track harmony learning in generative models.
To address these issues, we propose BandCondiNet, a conditional model based on parallel Transformers, designed to process the multiple music sequences and generate high-quality multitrack samples. Specifically, we propose multi-view features across time and instruments as high-fidelity conditions. Moreover, we propose two specialized modules for BandCondiNet: Structure Enhanced Attention (SEA) to strengthen the musical structure, and Cross-Track Transformer (CTT) to enhance inter-track harmony.
We conducted both objective and subjective evaluations on two popular music datasets with different sequence lengths. Objective results on the shorter dataset show that BandCondiNet outperforms other conditional models in 9 out of 10 metrics related to \emph{fidelity} and \emph{inference speed}, with the exception of Chord Accuracy. On the longer dataset, BandCondiNet surpasses all conditional models across all 10 metrics.
Subjective evaluations across four criteria reveal that BandCondiNet trained on the shorter dataset performs best in Richness and performs comparably to state-of-the-art models in the other three criteria, while significantly outperforming them across all criteria when trained on the longer dataset.
To further expand the application scope of BandCondiNet, future work should focus on developing an advanced conditional model capable of adapting to more user-friendly input conditions and supporting flexible instrumentation.

\end{abstract}



\begin{keyword}
Symbolic Music Generation; Conditional Music Generation; Multitrack Music Generation; Parallel Transformers; Multi-View Features.
\end{keyword}

\end{frontmatter}



\section{Introduction}
\label{sec_introduction}
Popular music, defined in this work as contemporary mainstream commercial music (e.g., pop, rock, and electronic) characterized by easily singable melodies and repetitive structures, serves as a vital form of art with a broad audience \citep{belkin2018musical}.
Recent advancements in generative artificial intelligence have enabled the development of sophisticated music generation systems \citep{herremans2017functional, civit2022systematic}. These innovations aim to make popular music creation more accessible, efficient, and personalized, even for individuals without formal music training. A key function of these systems is to empower users to create music tailored to specific conditions \citep{2020comprehensive}. For instance, users can customize their music by setting some simple parameters such as tempo and instrumentation \citep{von2023figaro}, or they can generate covers that are inspired from their favorite songs \citep{choi2023pop2piano}.

Various deep generative models have recently been proposed for the conditional generation of symbolic music. Global and fine-grained conditions are two main forms of control signals imbued in these conditional generative models. Global conditions typically use high-level music descriptions or abstract labels such as musical genres \citep{mukherjee2022composeinstyle, zukowski2023gtr}. Fine-grained conditions are achieved as a set of features that vary over time \citep{wu2023musemorphose, von2023figaro} or across instruments \citep{ens2020mmm, guo2022musiac}, capturing more detailed musical attributes and providing more powerful inductive bias than global conditions. On the other hand, popular music with a band setup is a type of multitrack music. Current multitrack music generation algorithms can be broadly categorized into image-based and sequence-based models, depending on how they represent the music \citep{le2025natural}. Image-based models treat multitrack music as multi-channel images, employing architectures similar to those used in image generation \citep{dong2018musegan, lin2024multi}. In contrast, sequence-based models represent multitrack music as single or multiple structured discrete sequences, processing them using approaches similar to those in language models \citep{Liu2022symphonynet, yu2022museformer, dong2023multitrack}. 

Despite significant advancements in conditional music generation and multitrack music generation, building a conditional generative system for multitrack popular music, the type of music with massive musical elements interweaving across time and instruments, still presents three key challenges: \textbf{1) developing high-fidelity features that convey rich musical information}, \textbf{2) improving the structural modeling of popular songs}, and \textbf{3) facilitating harmony learning across different tracks}.

Firstly, the fidelity of current conditional inputs for generative systems remains insufficient for multitrack music. Global conditions reflect music attributes only from a macro perspective, offering limited detail about entire songs. In contrast, current fine-grained conditions for multitrack music, which can vary either by track or over time, offer more precise control. However, track-differed features are calculated over the entire time span but still serving a global conditioning for each instrument. Time-varying features, which are calculated averagely at various time scales (e.g. at bar levels), may lack specificity for individual tracks, since multitrack music often contains diverse musical elements across different tracks.

\begin{figure}[t]
\centering
\includegraphics[width=5.4in]{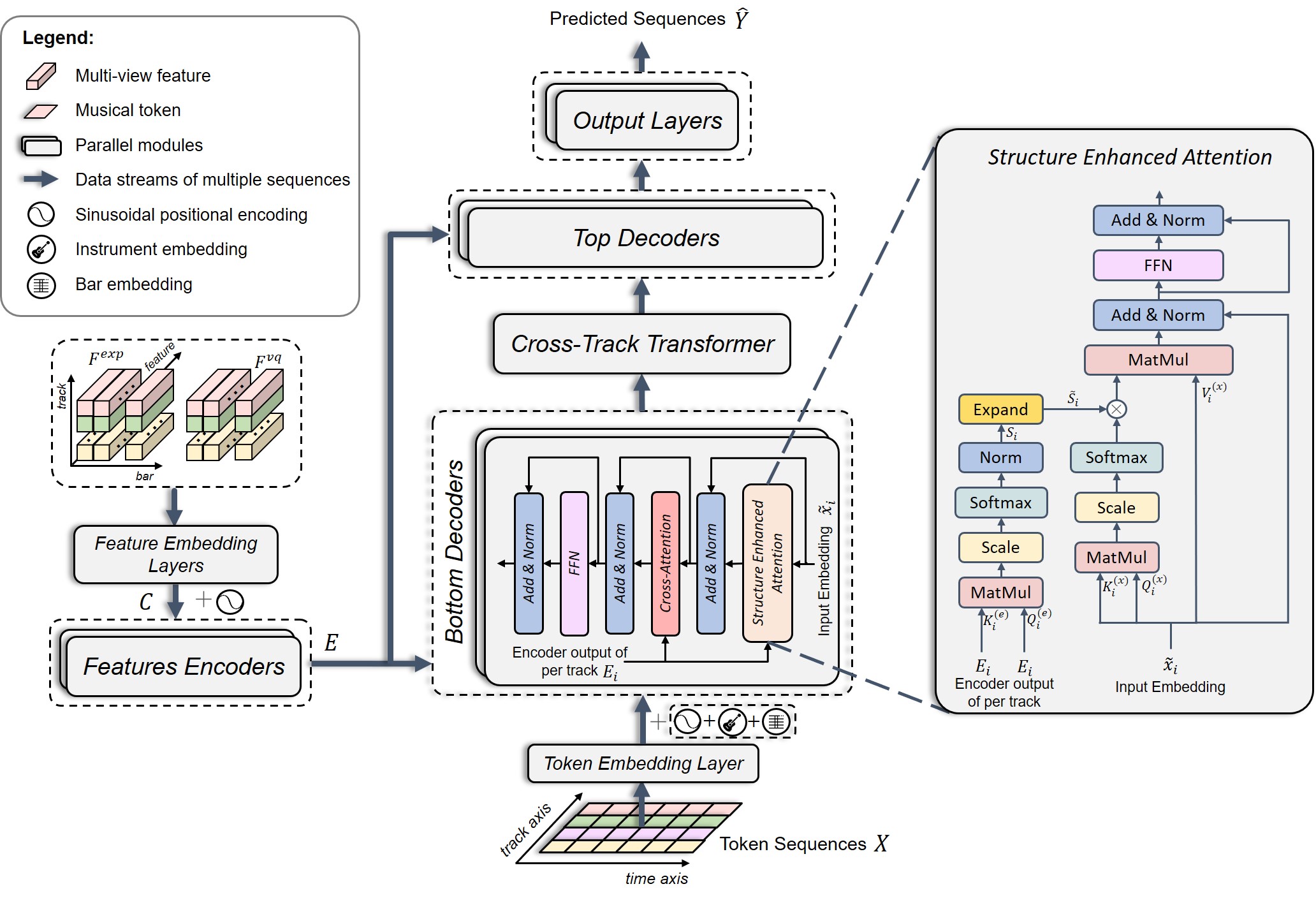}
\caption{The overall architecture of BandCondiNet. BandCondiNet is designed based on the canonical Sequence-to-Sequence framework and contains Feature Encoders, Bottom Decoders, Cross-Track Transformer (CTT), and Top Decoders. The Structure Enhanced Attention (SEA) module is utilized both in Bottom Decoders and the Top Decoders.}
\label{model_overall}
\end{figure}

Secondly, the structural modeling of popular music is underexplored in existing multitrack generation models. Popular songs often feature distinct structures with temporal dependencies across various time scales. These dependencies govern how musical elements unfold over time, linking individual notes into bars, phrases, and sections, thus maintaining the overall song structure. While previous studies have introduced various models for generating popular melodies \citep{zou2022melons} and single-track music \citep{zhang2022structure, wang2024whole}, only a few unconditional models, such as BandNet \citep{Zhou2019BandNet} and Museformer \citep{yu2022museformer}, specifically address the generation of multitrack popular music with a structured approach.

Thirdly, inter-track harmony in multitrack popular music, which defines how notes (e.g., pitch and rhythm patterns) from different tracks harmonize at any given moment \citep{ren2020popmag}, remains a significant challenge. Image-based models \citep{dong2018musegan, ding2023museflow} typically focus on overall harmony but overlook the local texture between tracks. Most sequence-based models \citep{dong2023multitrack, von2023figaro} tackle inter-track harmony modeling implicitly, relying solely on the naive generative model to capture interactions between notes across tracks. Some sequence-based models \citep{jin2022transformer, tie2024hybrid} incorporate specially designed modules (e.g. track-wise cross-attention mechanism) for harmony modeling, but these extra modules substantially increases both model complexity and the number of parameters.

To address the challenges outlined above, we propose BandCondiNet (as shown in Fig. \ref{model_overall}), a novel conditional music generation model built on a parallel Transformer framework, where individual Transformers handle the features or token sequences for each instrument. To tackle the first challenge of developing high-fidelity features, we introduce multi-view features as input conditions, extending the commonly adopted global and fine-grained conditions into three-dimensional feature matrix. The multi-view features provide a strong inductive bias from both the bar-view (per bar) and track-view (per track), significantly improving the fidelity of musical samples generated by BandCondiNet. During the encoding phase of BandCondiNet, a set of parallel Transformer encoders processes these multi-view features.

Furthermore, we represent multitrack music as multiple token sequences. Each sequence is tokenized using a Revamped MIDI(REMI)-like representation \citep{huang2020pop}, where each sequence corresponds to the notes of an individual instrument. These musical token sequences serve as inputs to the decoder block of BandCondiNet. To effectively fuse the multi-view conditions and enhance the overall quality of the generated music, BandCondiNet employs two parallel Transformer-based decoders, the Bottom Decoders and the Top Decoders, along with a Cross Track Transformer (CTT). Specifically, to address the second challenge of improving structural modeling, we design a novel Structure Enhanced Attention (SEA) module within Bottom Decoders and Top Decoders. SEA modules utilize multi-view features as comprehensive summary information at the bar level within each track, enabling adaptive learning of interrelations among different bars. Additionally, to tackle the third challenge of facilitating harmony learning across different tracks, the CTT module is proposed to facilitates cross-track interactions by introducing an additional attention computation for bar-related tokens, which ensures harmony and cohesion between tracks.

We conducted thorough experiments on two popular music datasets of different lengths to evaluate our proposed methods. 
For the objective evaluation, we introduce a total of 10 objective metrics, including 8 metrics related to \emph{fidelity} and 2 metrics focused on \emph{inference speed}. The experimental results on the shorter dataset show that BandCondiNet outperforms other conditional models across 9 of the 10 metrics, with the exception of Chord Accuracy. When trained on the longer dataset, BandCondiNet outperforms all conditional models across all 10 metrics. Notably, BandCondiNet exhibits minimal performance degradation across most metrics when compared to the results from the shorter dataset, highlighting its robustness in modeling long sequences.
For subjective evaluation, listening tests were conducted using four criteria: Coherency, Richness, Arrangement, and Overall Quality. Subjective results indicate that BandCondiNet trained on the shorter dataset excels in Richness and is comparable to state-of-the-art models in the remaining criteria, but its performance on a longer dataset significantly surpasses all benchmarks.

The remainder of this paper is as follows: Section~\ref{2_related_work} provides a comprehensive literature review on conditional music generation and multitrack music generation. Our proposed multi-view features and detail architecture of BandCondiNet are described in Section~\ref{3_methods}. Section~\ref{4_settings} and Section~\ref{5_results} present the experimental settings and results. Finally, the conclusion and discussion are drawn in Section~\ref{6_conclusion}.

\section{Related Work}
\label{2_related_work}
In this section, we provide a brief overview of symbolic music generation in terms of conditional music generation and multitrack music generation.

\subsection{Conditional Music Generation}
In recent studies on symbolic music generation, various types of conditions have been proposed to guide the creative process. These conditions can be derived from diverse modalities, such as images \citep{wang2023continuous}, video \citep{kang2024video2music}, audio \citep{choi2023pop2piano}, and text \citep{zhang2023controllable, lu2023musecoco}, enabling interaction of abstract semantic relationships between music and other media. However, a more common and effective approach to conditional music generation involves using conditions derived from musical features and elements. These conditions provide generative models with musical domain-specific knowledge. This paper primarily focuses on such musically driven conditions. Based on their level of granularity, these conditions can be broadly categorized into two types: global conditions and fine-grained conditions, as highlighted in Table~\ref{related_work_comparison}. Both types are discussed in detail below. 

\begin{table}[!t]
\tiny
\caption{Summary of music types (Multi-Track vs. Other) and music conditions (Global vs. Fine-Grained) in existing studies. Note that fine-grained conditions can also imply global conditions, as a global condition can be replicated by fixing a conditioning attribute to a constant value across all time steps.}
\label{related_work_comparison}
\centering
\begin{tabular}{lcccc}
\toprule
\multirow{2}{*}{\textbf{Method}} & \multirow{2}{*}{\textbf{Multi-Track}} & \multirow{2}{*}{\textbf{Global Conditions}} & \multicolumn{2}{c}{\textbf{Fine-Grained Conditions}}  \\
\cmidrule(lr){4-5} &  & & \textbf{Bar-wise} & \textbf{Track-wise}  \\ \midrule
MuseNet~\citep{payne2019musenet}         & \usym{2713} & \usym{2713}  & \usym{2717} & \usym{2717}  \\
GTR-CTRL~\citep{zukowski2023gtr}  & \usym{2717} & \usym{2713} & \usym{2717} & \usym{2717} \\
MMM~\citep{ens2020mmm} & \usym{2713} &  \usym{2713} & \usym{2717} & \usym{2713} \\
MusIAC~\citep{guo2022musiac} & \usym{2713} & \usym{2713} & \usym{2713} & \usym{2717} \\
MuseMorphose~\citep{wu2023musemorphose} & \usym{2717} & \usym{2713} & \usym{2713} & \usym{2717} \\
FIGARO~\citep{von2023figaro} & \usym{2713} & \usym{2713} & \usym{2713} & \usym{2717} \\
SCG~\citep{huang2024symbolic} & \usym{2717} & \usym{2713} & \usym{2713} & \usym{2717} \\
BandCondiNet(Ours) & \usym{2713} & \usym{2713} & \usym{2713} & \usym{2713}\\
\bottomrule
\end{tabular}
\end{table}

Global conditions are widely used in conditional music generation. In some studies, global condition is encoded into parts of music representation and serves as prompts to guide the creation process \citep{payne2019musenet, zukowski2023gtr}. Other research treats global conditions as supervision signals, incorporating them into the training procedure \citep{tan2020music, luo2020mg, bao2023generating}. However, global conditions have some limitations: for instance, the generative models often struggle to retain these conditions, especially when tasked with generating complex music or long sequences, where the influence of global conditions tends to diminish over time.

In addition to global conditions, several studies have explored fine-grained conditions \citep{ens2020mmm, guo2022musiac} that vary across instruments or over time. Predefined music descriptors \citep{wu2023musemorphose, huang2024symbolic} such as chord progressions, note density, polyphony rate, rhythmic intensity, are commonly used input conditions. Alternatively, Musical features can be driven from deep neural networks through representation learning \citep{choi2020encoding, von2023figaro}. The abovementioned fine-grained conditions are typically extracted at the track or bar level, but each captures only a partial perspective of the attributes in multitrack music. Track-level features summarize the characteristics of individual tracks but provide only track-wise differences, while bar-level features capture temporal variations but neglect the nuance between different tracks.

In this paper, we address these limitations by introducing multi-view features, which provide generative models with high-fidelity musical information at both the track and bar level. It is important to distinguish our work from the recently proposed Multi-view MidiVAE \citep{lin2024multi}, which also employs the term ``multi-view''. The core concept in their work pertains to a novel hybrid VAE-based architecture, featuring distinct track-view and bar-view VAE. In contrast, the ``multi-view'' aspect of our method lies not in the model's architecture, but in the explicitly designed, high-fidelity features that we use as external conditions to guide the generation process.

\subsection{Multitrack Music Generation}
Creating popular music with a band setup falls under the task of multitrack music generation task. Multitrack music can be visualized as multi-channel images, several studies approach it as an image generation problem. They first convert each track into a pianoroll or its advanced versions, and then apply various generative models to reconstruct each pixel (i.e., note) of the pianoroll \citep{dong2018musegan, ding2023museflow, zhao2023accomontage}. 

Alternatively, symbolic music could also be represented as sequences of structured discrete tokens. Previous work \citep{ens2020mmm, von2023figaro} typically merge notes from various tracks into a single, lengthy sequence. To reduce the sequence length in multitrack music, some approaches employ techniques such as tuple encapsulation \citep{ren2020popmag, zeng2021musicbert, dong2023multitrack}, token combinations \citep{Donahue2019LakhNES, liang2020pirhdy}, and data compression methods \citep{Liu2022symphonynet, fradet2023byte}. Additionally, a few studies \citep{jin2022transformer, tie2024hybrid} encode each track individually, transforming multitrack music into multiple sequences rather than merging them into one. These methods often utilize parallel recurrent neural networks (RNNs) or Transformers to generate multitrack sequences concurrently.

Popular music is known for its emotional resonance and widespread appeal, which are partly achieved through well-defined structures and harmonically engaging compositions. These two core elements, structure and inter-track harmony, are crucial in the multitrack popular music generation process. Structure corresponds to temporal dependencies across varying time scales, while inter-track harmony involves polyphonic interactions among notes from different tracks. Despite their importance, only a few unconditional models, such as BandNet \citep{Zhou2019BandNet} and Museformer \citep{yu2022museformer}, specifically address the challenge of generating multitrack music with recognizable structures. BandNet uses predefined templates as structural constraints, while Museformer employs a fine-grained attention mechanism based on the assumption that popular music structures consist of replicated bars separated by fixed intervals. However, both models rely heavily on predefined templates or prior assumptions, which limit their ability to adaptively learn relationships between musical bars.

In terms of inter-track harmony modeling, many image-based models \citep{dong2018musegan, ding2023museflow} enhance harmony between tracks by sharing a common latent vector or fusing latent vectors from each track. While this method maintains overall harmony, it frequently overlooks local textures between tracks, which are essential for detailed musical coherence. In contrast, sequence-based models typically merge notes from different tracks into a single sequence and directly feed the them into sequential models. This approach handles implicitly inter-track harmony but tends to disrupt the intrinsic harmony and texture of multitrack music. Some models using parallel RNNs and Transformers \citep{jin2022transformer, tie2024hybrid} introduce cross-attention mechanisms to integrate note-level information between every two tracks. However, as the number of musical tracks increases, the number of required cross-attention modules for processing all track pairs grows substantially, resulting in significantly increased model complexity and computational demands.

To address the aforementioned challenges related to structure modeling and inter-track harmony, we introduce the Structure Enhanced Attention (SEA) module and the Cross-Track Transformer (CTT) as integral components of the BandCondiNet decoder. The detailed architecture and functionality of these components will be presented in the next section.

\section{Methods}
\label{3_methods}
In this section, we first introduce two sets of multi-view features as three-dimensional high-fidelity conditions, which automatically extracted from multitrack music at both the bar and track levels. Then, the architecture of BandCondiNet, whose overall structure was previously shown in Fig.~\ref{model_overall}, is presented.

\subsection{Multi-view Features}
\begin{figure}[t]
\centering
\includegraphics[width=5.4in]{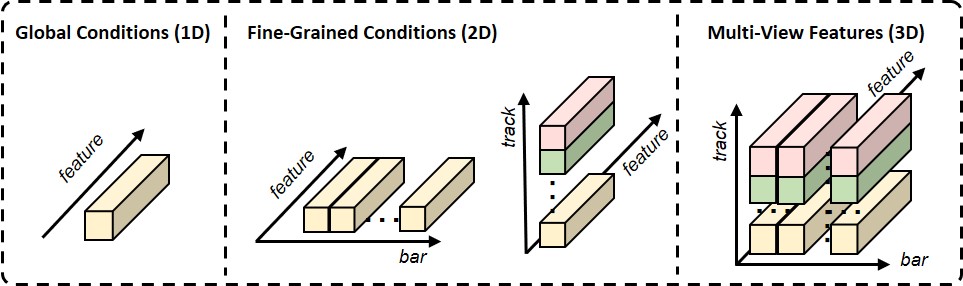}
\caption{A visual comparison of various conditions used in conditional music generation: on the left are global conditions, in the middle are the commonly adopted fine-grained conditions, and on the right are our proposed multi-view features.}
\label{multiview_features}
\end{figure}

To provide high-fidelity features that deliver rich musical information and effectively guide the generation of multitrack music, we introduce multi-view features. These features are computed on a bar-by-bar basis over time and on a track-by-track basis across instruments. Formally, the multi-view features are stored in a three-dimensional matrix $F\in \mathbb{R}^{I\times B\times d_f}$, where $I$ represents the instrument (i.e., track) numbers, $B$ represents the bar numbers, and $d_f$ is the dimension of the feature. As illustrated in Fig.~\ref{multiview_features}, compared to global conditions (which typically consist of only a single dimension) and other fine-grained conditions (which usually span two dimensions, varying either over time or across tracks), multi-view features contain richer, high-fidelity musical information. This allows generative models to benefit from a strong inductive bias, drawing on both track-level and bar-level perspectives. In this paper, we extend the two-dimensional expert and learned features initially defined by the FIGARO \citep{von2023figaro} into three-dimensional multi-view features, with several key adjustments, as discussed below.

\subsubsection{Expert Features}
\label{3_1_1_expert_features}

Expert features are achieved by deriving a set of high-level musical attributes, grounded in domain knowledge, that are compositionally meaningful and intuitive to experts. Specifically, our selected seven features include chord types, drum types, drum density, note density, and the mean values of pitch, velocity, and duration. These attributes are calculated from each bar and each track and their selection aligns with established practices in many conditional music generation tasks \citep{wu2023musemorphose, von2023figaro}. Here, we exclude musical attributes as expert features like time and key signatures, as they typically remain static across bars and tracks in most popular music and do not offer the bar-by-bar and track-by-track dynamic control.

\begin{itemize}
\item Chord Types (CT): common chords extracted with a chord detection tool\footnote{\url{https://github.com/joshuachang2311/chorder}} per beat, such as G half-diminished seventh chord, which is written as [Chord\_G:hd7].
\item Drum Types (DT): total number of drum types. 
\item Drum Density (DD): the number of drum note onsets per beat.
\item Note Density (ND): similar to drum note density but calculated on the non-drum notes set.
\item Mean Pitch (MP), Mean Duration (MD), and Mean Velocity (MV): the average pitch, duration, and velocity of non-drum notes, respectively.
\end{itemize}

All multi-view expert features $F^{exp} \in \mathbb{R}^{I\times B\times d_f}$ excluding CT are extracted bar-wise on every track. CT is a 4-element tuple $\{\mathrm{CT}^1_b, \mathrm{CT}^2_b, \mathrm{CT}^3_b, \mathrm{CT}^4_b\}$ where each element is inferred per beat during the given bar (only consider the time signature of $4/4$).   The expert features $f^{exp}_{i,b}\in \mathbb{R}^{d_f}$ at $i^{th}$ track and $b^{th}$ bar can be written as:
\begin{small}
\begin{equation}
\label{expert_feature}
\begin{aligned}
f_{i,b}^{exp}= & \left\{\begin{array}{ll}
\{\mathrm{DT}_{i,b}, \mathrm{DD}_{i,b}\} & \text{for } i \text { is drum;} \\
\{\mathrm{ND}_{i,b}, \mathrm{MP}_{i,b}, \mathrm{MD}_{i,b}, \mathrm{MV}_{i,b}\} \oplus \mathrm{CT}_b & \text {otherwise. }
\end{array} \right. \\
\mathrm{CT}_b= & \{\mathrm{CT}^1_b, \mathrm{CT}^2_b, \mathrm{CT}^3_b, \mathrm{CT}^4_b\}
\end{aligned}
\end{equation}
\end{small}
where $\oplus$ is a concatenation operator for two tuples. The expert features of the drum track are organized as a 2-element tuples (i.e., $d_f=2$), while that of other instruments contains eight features (i.e., $d_f=8$) and the 4-element CT is shared between instruments.

\subsubsection{Learned Features}
In recent studies on image generation \citep{esser2021taming} and music generation \citep{wang2023continuous, copet2024simple}, the Vector Quantized Variational AutoEncoder (VQ-VAE) \citep{van2017neural} is used to learn powerful and robust representations for generating high-quality samples. Inspired by these studies and FIGARO \citep{von2023figaro}, we consider that learned features can be extracted from the latent space of a pre-trained VQ-VAE model. Unlike FIGARO, however, we extract multi-view learned features from the bar-level token sequence of each individual track, rather than from a merged sequence that covers all tracks.

The VQ-VAE model mainly consists of an encoder block, a decoder block, and a vector quantization (VQ) block. The VAE encoder first maps the bar-level token sequence of an individual track to the latent space. For a richer latent representation, we use the decomposed vector quantization scheme \citep{kaiser2018fast} to slice the latent dimension of encoder output $z_e\in\mathbb{R}^{d_l}$ into 8 groups $z_e=\{z^1_e, z^2_e, \cdots, z^8_e\}$, where $z^i_e\in \mathbb{R}^{d_l/8}$, $d_l$ denotes the initial latent dimension before slicing operation. Then, 8 groups of VQ codes $z^{vq}=\{z^{vq}_n\}^8_{n=1} \subset \mathbb{R}^K$ are produced with a shared codebook of size $K$ after the VQ block. These codes are finally fed to Transformer decoder to reconstruct the original musical sequence. Such the 8-code grouping is considered as the learned features $F^{vq} \in \mathbb{R}^{I\times B\times d_f}$ as follows.
\begin{equation}
\label{learned_feature}
   f_{i,b}^{vq}=\{z^{vq}_{i,b,n}\}^8_{n=1}
\end{equation}
where $f_{i,b}^{vq} \in \mathbb{R}^{d_f}$ is the entry of $F^{vq}$ at $i^{th}$ track and $b^{th}$ bar and is the set of all eight VQ codes (i.e., $d_f=8$).

\subsection{Model Architecture}
We utilize a parallel Transformers-based architecture as the backbone of BandCondiNet for conditional music generation, where individual Transformers process both the multi-view features and musical token sequences for each instrument. As shown in Fig.~\ref{model_overall}, BandCondiNet comprises two main blocks, namely the feature encoders block and the decoders block. In the encoding phase, the three-dimensional multi-view features are processed by a set of parallel Transformer encoders, each independently handling the high-fidelity conditions for its corresponding track. During the decoding phase, we introduce three decoding modules: Bottom Decoders, the Cross-Track Transformer (CTT), and Top Decoders. The Structure Enhanced Attention (SEA) module, integrated into both the Bottom and Top Decoders, improves structural modeling, while the CTT module enhances inter-track harmony modeling.

\subsubsection{Feature Encoders}
We first convert $F^{exp}$ and $F^{vq}$ into an embedding vector. Given an expert feature $f^{exp}_{i,b} \in \mathbb{R}^{d_f}$ extracted from the $i^{th}$ track and $b^{th}$ bar, as defined in Eq.~\eqref{expert_feature}. We embed each component feature $f^{exp}_{i,b, m}$ and concatenate them to obtain their embedding vector ${C}^{exp}_{i,b}$, where $m$ represent a specific type of music features within set $\mathcal{M}$. We can obtain the embedding vector ${C}^{vq}_{i,b}$ of the learned features in a similar way. Then, we combine the two embedding vectors to get ${C}_{i,b} \in \mathbb{R}^d$, where $d$ is the embedding dimension. The above computational procedure can be formulated as follows.

\begin{equation}
\label{eq_embeddings}
    \begin{aligned}
        &{C}^{exp}_{i,b}=\mathrm{Concat}([\mathrm{FE}^{exp}_m({f^{exp}_{i,b, m}}), m\in \mathcal{M}]) \\
        &{C}^{vq}_{i,b}=\mathrm{Concat}([\mathrm{FE}^{vq}({f^{vq}_{i,b, n}}), n=1,\cdots,8]) \\
        &{C}_{i,b}=\mathrm{Concat}([{C}^{exp}_{i,b}, {C}^{vq}_{i,b}])
    \end{aligned}
\end{equation}
where $\mathrm{Concat}$ denotes the vector concatenation operation, $\mathrm{FE}^{exp}_m$ represents the embedding layer for the expert feature of the type $m$, and $\mathrm{FE}^{vq}$ is the shared embedding layer for the learned feature of every group $n$.


After embedding, we get $C=\{\{C_{i,b}\}^B_{b=1}\}^I_{i=1} \in \mathbb{R}^{I \times B \times d}$ as the input fed to the parallel feature encoders. Each encoder $\mathrm{Enc}_i$ aims to handle the individual bar-varying vectors $\{C_{i,b}\}^B_{b=1}$, and consists of a vanilla Transformer encoder. Here, we add the bar index $b$ as the position information into the vanilla \emph{sinusoidal} position encoding $\mathrm{PE}$. The final encoder output ${E}_{i,b}$ corresponding to the input $C_{i,b}$ can be obtained by:
\begin{equation}
\label{eq_enc}
    {E}_{i,b}=\mathrm{Enc}_i(C_{i,b} + \mathrm{PE}(b))
\end{equation}

\subsubsection{Bottom Decoders}
To align with the parallel Transformers architecture, multitrack popular music is tokenized as multiple token sequences $X=\{\{x_{i,t}\}^T_{t=1}\}^I_{i=1}\in \mathbb{R}^{I \times T}$, where each musical sequence $\{x_{i,t}\}^T_{t=1}$ follows a REMI-like representation (detailed in Section~\ref{4_2_remi_like}) corresponds to a specific instrument. Here, $T$ denotes the maximum length among all musical token sequences. We then obtain the embedded input sequences $\tilde{X}=\{\{\tilde{x}_{i,t}\}^T_{t=1}\}^I_{i=1}\in \mathbb{R}^{I \times T \times d}$ as:
\begin{equation}
\label{music_embedding}
    \tilde{x}_{i,t} = \mathrm{TE}({x_{i,t}}) + \mathrm{PE}(t) + \mathrm{BE}(b) + \mathrm{IE}(i)
\end{equation}
where $\mathrm{TE}$, $\mathrm{BE}$, and $\mathrm{IE}$ are the embedding layers for the token, bar index, and instrument types respectively. Then, we utilize $I$ parallel Transformer decoders to constitute the Bottom Decoders, each dedicated to modeling the relationships between tokens within an individual track autoregressively. The architecture of each Bottom Decoder $\mathrm{Dec}_i^{btm}$ is similar to that of a vanilla Transformer decoder, but with one key difference: the original self-attention module \citep{vaswani2017attention} is replaced by our proposed Structure Enhanced Attention (SEA) module. 

\begin{figure}[!t]
\centering
\includegraphics[width=4.6in]{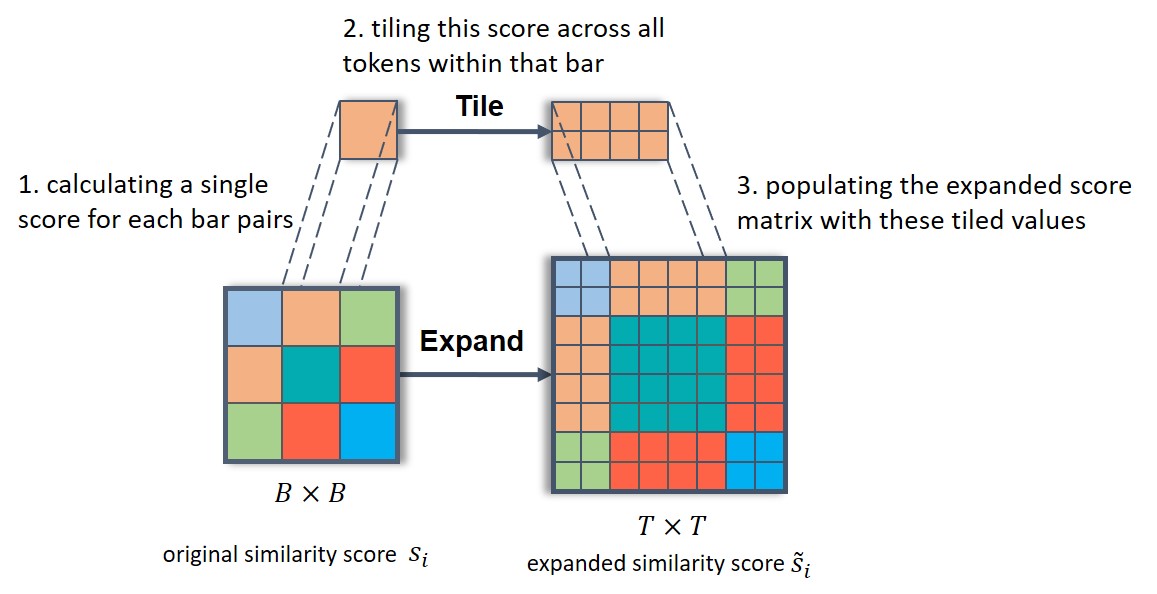}
\caption{The expansion operation of the SE-SA module consists of three steps: 1) calculating a single score for each bar pairs, 2) tiling this score across all tokens within that bar, and 3) populating the expanded score matrix with these tiled values. $B$ denotes the bar numbers, and $T$ represents the length of token sequences.}
\label{expand_module}
\end{figure}

The SEA module is built on the assumption that the multi-view features provide sufficient summary information to characterize notes at the bar-level scope within a track. Hence, we could learn internal interrelations among various bars from the encoded multi-view features $E_i=\{E_{i,b}\}^B_{b=1}$, and then use these learned interrelations to guide the structure modeling of the entire sequence for each track. 

Concretely, we first compute the similarity score $S_i$ between different bars for each track as follows:

\begin{equation}
\label{eq_sim_matrix}
    S_i = \mathrm{LayerNorm}(\mathrm{Softmax}(\frac{Q^{(e)}_i {K^{(e)}_i}^T}{\sqrt{d}}))
\end{equation}
where $Q^{(e)}_i=W^{(e)}_Q E_i$, $K^{(e)}_i=W^{(e)}_K E_i$, and $W^{(e)}_Q$, $W^{(e)}_K$ are trainable parameters to project $E_i$ to queries and keys respectively. Subsequently, as Fig.~\ref{expand_module} shows, we expand the matrix $S_i\in \mathbb{R}^{B\times B}$ of each track to a larger $\tilde{S}_i\in \mathbb{R}^{T\times T}$ by tiling the similarity score of a bar to all tokens belonging to this bar. Then, we define the computation of our proposed SEA for $\tilde{x}_i=\{\tilde{x}_{i,t}\}^T_{t=1}$ in $\tilde{X}$ as:
\begin{equation}
\label{eq_sea}
    Attn(\tilde{x}_i, \tilde{S}_i) = \mathrm{Softmax}(\frac{\tilde{S}_i \otimes (Q^{(x)}_i {K^{(x)}_i}^T)}{\sqrt{d}}) V^{(x)}_i
\end{equation}
where $Q^{(x)}_i=W^{(x)}_Q \tilde{x}_i,$, $K^{(x)}_i=W^{(x)}_K \tilde{x}_i$, $V^{(x)}_i=W^{(x)}_V \tilde{x}_i$, and $W^{(x)}_Q$, $W^{(x)}_K$, $W^{(x)}_V$ are trainable parameters; The symbol $\otimes$ denotes the element-wise multiplication operation. For simplicity, we omit the computation of multi-head concatenation here. 

The cross-attention modules in Bottom Decoders are kept the same as the vanilla Transformer decoder to fuse the encoded multi-view features $E_i=\{E_{i,b}\}^B_{b=1}$ of each track. After the parallel Bottom Decoders with the SEA module, we obtain the output ${O}^{btm}_{i,t}$ of each token as:
\begin{equation}
\label{btm_dec}
{O}^{btm}_{i,t} = \mathrm{Dec}_i^{btm}(\tilde{x}_{i,t}, {E}_{i})
\end{equation}

\begin{figure}[!t]
\centering
\includegraphics[width=5.2in]{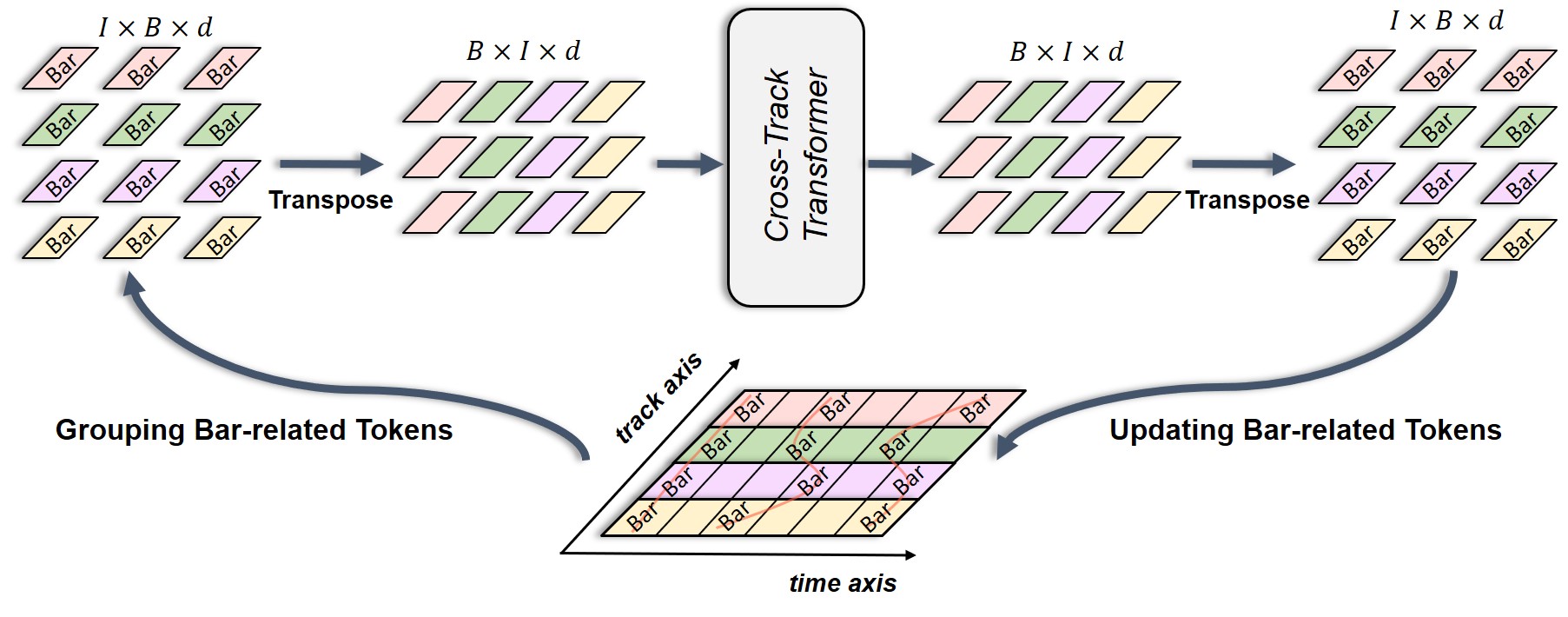}
\caption{Grouping and Updating operations of Cross-Track Transformer. $B$ denotes the bar numbers, $I$ denotes track numbers, and $d$ represents the hidden size.}
\label{cross_transformer}
\end{figure}

\subsubsection{Cross-Track Transformer}
Our parallel Bottom Decoders operate independently, each focusing solely on a specific track. This independence leads to a loss of inter-track harmony among various instruments. To address this situation, we propose a novel Cross-Track Transformer (CTT) to construct the connections between different tracks by enhancing the attention of bar-related tokens.

First, we group the bar-related tokens $O^{btm}_{i,t_b}$ from all tokens set $\{O^{btm}_{i,t}\}^T_{t=1}$ within every track when the $b^{th}$ bar-related token is located at position $t_b$. Given all bar-related tokens set $\{\{{O}^{btm}_{i,t_b}\}^B_{b=1}\}^I_{i=1} \in \mathbb{R}^{I\times  B \times d}$ after all grouping operations, we subsequently transpose them to a new set $\{\{{O}^{btm}_{i,t_b}\}^I_{i=1}\}^B_{b=1} \in \mathbb{R}^{B\times  I \times d}$, where the instrument axis is considered as the sequential dimension. Later, we use a vanilla Transformer encoder $\mathrm{Enc}^{ctt}$ to learn the relationships between bar tokens from various instruments as:
\begin{equation}
{O}^{crs}_{i,t_b} = \mathrm{Enc}^{ctt}({O}^{btm}_{i,t_b})
\end{equation}
Finally, the output ${O}^{crs}_{i,t_b}$ of $\mathrm{Enc}^{ctt}$ is restored to the same shape as the original $O^{btm}_{i,t_b}$. In addition,  the final output ${\tilde{O}}^{btm}_{i,t}$ is obtained by updating bar-related tokens as follows.
\begin{equation}
\label{eq_update}
{\tilde{O}}^{btm}_{i,t}=\left\{\begin{array}{ll}
{O}^{crs}_{i,t_b} & t=t_b, \\
{O}^{btm}_{i,t} & \text { otherwise. }
\end{array}\right.
\end{equation}
For better comprehension, the visual procedure of CTT is shown in Fig.~\ref{cross_transformer}. 

\subsubsection{Top Decoders}
The architecture of Top Decoders is the same as that of the Bottom Decoders, and equally includes the SEA module. Similar to the Eq.~\eqref{btm_dec}, given each top decoder $\mathrm{Dec}_i^{top}$ and encoded conditions ${E}_{i}$ of each track, we obtain the output ${O}^{top}_{i,t}$ as:
\begin{equation}
{O}^{top}_{i,t} = \mathrm{Dec}_i^{top}({\tilde{O}}^{btm}_{i,t}, {E}_{i})
\end{equation}

\begin{algorithm}
\caption{Training Procedure of BandCondiNet.}
\label{algorithm_1}
\begin{algorithmic}[1]
\State \textbf{Input:} Multi-view features $F^{exp}$, $F^{vq}$; music sequences $X$
\State \textbf{Output:} Optimal parameters of BandCondiNet
\State \textbf{Initialize:} Model parameters (encoders, decoders, embeddings of BandCondiNet)
\While{not converged}
    \For{each batch}
        \State Encode multi-view features $F$ by Feature Encoders
        \State Prepare music embeddings $\tilde{X}$
        \State Compute and expand similarity matrix $S_i$
        \State Process musical embeddings by Bottom Decoders to obtain $O^{btm}$
        \State Update bar-level information using Cross-Track Transformer 
        \State Process refined representations ${\tilde{O}}^{btm}$ using Top Decoders similar to Bottom Decoders
        \State Generate predicted tokens $\hat{Y}$ from output $O^{top}$ of Top Decoders
        \State Compute training loss $\mathcal{L}$
        \State Backpropagate and update parameters
    \EndFor
\EndWhile
\end{algorithmic}
\end{algorithm}

\subsubsection{Training and Inference}

Considering that the distribution of notes on each track varies, we adapt an independent linear layer $\mathrm{MLP}_i$ for the $i^{th}$ track to obtain the prediction value $\hat{y}_{i,t}$ from the output ${O}^{top}_{i,t}$ as:

\begin{equation}
\label{eq_output}
\hat{y}_{i,t} = \mathrm{Softmax}(\mathrm{MLP}_i({O}^{top}_{i,t}))
\end{equation}

In training, let $Y$ be the ground truth, which is the right-shifted version of input sequences $X=\{\{x_{i,t}\}^T_{t=1}\}^I_{i=1}$. The training objective of BandCondiNet can be formulated as the reconstruction loss of multiple sequences between $Y$ and  $\hat{Y}=\{\{\hat{y}_{i,t}\}^T_{t=1}\}^I_{i=1}$ as:
\begin{equation}
\mathcal{L}(Y, \hat{Y})=\sum_{i=1}^I \sum_{t=1}^T \mathrm{CELoss}\left(y_{i,t}, \hat{y}_{i,t}\right)
\end{equation}
where $\mathrm{CELoss}$ denotes cross-entropy loss function. The overall training flow of BandCondiNet is shown in Algorithm~\ref{algorithm_1}.

During inference, we select a piece of music from the testing dataset as the reference and extract multi-view features $F$ from it. Then, we feed these conditions $F$ to the encoder block of BandCondiNet and feed a seeding sequence that begin with the indicator [BOS] for each track to the decoders block. Finally, each musical token is sampled from the predicted value $\hat{y}_{i,t}$. In addition, we adopt a top-$k$ sample strategy with the $k$ set to 2\% of the vocabulary size. The sampling process is stopped when the number of bars increases that of the reference.

\section{Experimental Setup}
\label{4_settings}
\subsection{Dataset}
We conduct our experiments using a subset of popular music from the LakhMIDI dataset (LMD) \citep{raffel2016learning}, which is the largest publicly available symbolic music dataset with multiple instruments. To ensure the quality of data, we perform several preprocessing steps following previous research \citep{dong2018musegan, ren2020popmag, yu2022museformer}, including genre selection, melody extraction, instrument compression, and data filtering. 

Firstly, we select the MIDIs in the LMD-matched subset that fall into five genres (i.e., \emph{Pop}, \emph{Rock}, \emph{Electronic}, \emph{Dance}, \emph{Country}), according to the metadata matched to the Million Song Dataset \citep{bertin2011million}. Secondly, MIDI Miner \citep{guo2019midi} is employed to identify the melody track. Thirdly, we compress all instruments into six types (i.e., Drum, Piano, Guitar, Bass, Strings, Square Synthesizer, with the last one playing the melody). Finally, we apply a set of filtering rules to remove low-quality songs as listed below. These rules are grounded in three guiding principles: (1) ensuring structural completeness, (2) avoiding unnecessary complexity, and (3) reducing duplicated musical content.
\begin{itemize}
    \item Remove the duplicated songs that have the same expert features defined in Section~\ref{3_1_1_expert_features}.
    \item Only keep songs with the time signature of \textbf{$4/4$}.
    \item Convert the BPM of all songs to 120.
    \item Only keep songs with at least 4 types of instrument (must contain a Square Synthesizer (Melody) and Drum, the remaining two tracks are selected with the largest number of notes).
    \item Only keep songs that have more than 16 bars and 512 notes.
    \item Remove songs that contain more than 4 empty bars.
    \item Transfer the tonality of songs into ``C major'' or ``A minor'' to mitigate key-induced bias in pitch modeling.
\end{itemize}

After completing the preprocessing, a total of 5,787 musical pieces remain in our final dataset. To simply the modeling process, we limit each song to a maximum of four tracks: Melody, Drums, and two additional accompaniment tracks. Due to the large variation in song lengths and the limitation of computing resources, we choose not to train our model on full-length songs. Instead, we convert the dataset into two versions: a 32-bar and a 64-bar datasets, to evaluate the model performance in generating both short-term and long-term sequences. 

For the 32-bar dataset, we segment each song using a flexible window \footnote{To determinate the flexible window size, we applied the phrase boundary detection algorithm \citep{dai2020automatic} to roughly recognize structure of each song. The window size is depended on the total bar numbers of several consecutive phrases.} of 16 to 32 bars with a constant stride of 8 bars. Similarly, for the 64-bar dataset, we apply a larger window ranging from 32 to 64 bars, using the same constant stride. The detailed statistics of both datasets are shown in Table~\ref{dataset_stats}. We randomly reserve 10\% of songs and their corresponding segments for testing, while the remaining data is used for model training.

\begin{table}[!t]
\scriptsize
\caption{The statistics of the 32-bar and 64-bar datasets.}
\label{dataset_stats}
\centering
\begin{tabular}{ccccc}
\toprule
\textbf{Datasets} & \textbf{\# Samples}& \textbf{\# Average Bars}& \textbf{\# Average Notes}& \textbf{Total Duration} \\
\midrule
32-bar & 46,004& 24.7 & 875.3 & 632h \\
64-bar & 36,527& 48.3 & 1,847.2 & 981h \\
\bottomrule
\end{tabular}
\end{table}

\subsection{Music Representation}
\label{4_2_remi_like}
The Revamped MIDI(REMI) representation \citep{huang2020pop} is a beat-based music tokenization method that provides an explicit metrical grid for represent symbolic music. REMI typically represents music using three note-related tokens (Pitch, Duration and Velocity) and four metric-related tokens (Bar, Position, Tempo and Chord). In this paper, we convert multitrack music into multiple token sequences to accommodate the parallel Transformer framework, where each sequence corresponds to a specific instrument and is tokenized in a REMI-like representation. To effectively represent the musical notes in each instrument and reduce the lengths of musical token sequences,  we introduce several modifications to the original REMI as follows.
\begin{itemize}
    \item We exclude the Tempo and Chord tokens, as the tempo values in our dataset are fixed during data preprocessing, and our proposed multi-view features already incorporate chord information.
    \item We introduce metric-related tokens to encode instrument information at the beginning of each token sequence. In line with our data preprocessing, we consider six common instrument types in popular music: Drum, Piano, Guitar, Bass, Strings, and Square Synthesizer (for melody).
    \item We use only two bar-related tokens (i.e., [Bar\_Empty] and [Bar\_Normal]), to indicate whether the current bar is empty or not.
    \item Unlike previous studies that often omit drum notes or represent them using shared pitch vocabularies from other instruments, we treat drum notes as an independent token (i.e., [Pitch\_D]) to indicate the drum type. To further reduce the sequence length for the drum track, we omit the duration and velocity tokens for drum notes, setting both to default values of a $16^{th}$ note duration and a velocity of 64, respectively.
    \item Considering that REMI-like sequences can become lengthy, especially for long pieces of music, we apply Byte Pair Encoding (BPE) \citep{fradet2023byte} to compress each track's sequence. Specifically, BPE is trained on a corpus of note-related tokens that share the same note onset position, while metric-related tokens are excluded from the learning process.
\end{itemize}

We use 128 pitch values for non-drum notes and 31 drum types. The temporal resolution is set to 48-time steps per quarter note. Note velocity and duration are quantized using the same methods as in the FIGARO model \citep{von2023figaro}. The raw vocabulary size of our REMI-like representation is 284, including three special tokens ([PAD] for padding, [BOS] for the beginning of the sequence, and [EOS] for the end). After applying BPE, the vocabulary size expands to 10,000.

\subsection{Benchmark Models}
\label{4_3_benchmark_models}
We compare our proposed BandCondiNet model with four benchmark models: two unconditional models, and the two state-of-the-art methods for conditional multitrack music generation. The benchmark models are as follows.
\begin{itemize}
    \item Music Transformer \citep{Huang2019musictransformer}: An unconditional Transformer-based model that uses a relative attention mechanism.
    \item BandCondiNet-base: Another unconditional baseline. It discards the feature encoders of BandCondiNet and utilize multiple Transformer decoders in parallel. Each decoder shares the same architecture as the decoder of Music Transformer and is responsible for generating notes in the corresponding instrument.
    \item MuseMorphose \citep{wu2023musemorphose}: A conditional model that pairs a Transformer with a Variational Autoencoder(VAE). It introduces two bar-level musical features namely rhythmic intensity and polyphony for fine-grained music generation, also supporting style transfer in latent space.
    \item FIGARO \citep{von2023figaro}: Another conditional model built on a Transformer-based framework. It uses bar-level (two-dimensional) expert features and learned features as input conditions. Our three-dimensional multi-view features approach extends this by adding track-level information and represents comprehensive musical information from every track and every bar.
\end{itemize}

Among these benchmark models, only BandCondiNet-base employs parallel Transformers as does our BandCondiNet. The rest of the benchmark models use a single Transformer.

\subsection{Model Configurations}
The encoder layers, decoder layers, attention heads, and hidden size of the feed-forward networks for all benchmark models (where applicable) are set to 4, 6, 8, and 1,024 respectively. Our proposed BandCondiNet maintains the same parameter settings as the benchmarks except that the number of layers in the Bottom Decoders, Top Decoders, and Cross-Track Transformer is set to 3, 3, and 2 respectively. For all models, the hidden size and embedding size are uniformly set to 256. Additionally, BandCondiNet utilizes four parallel feature encoders, Bottom Decoders, and Top Decoders to accommodate the fixed four tracks in our dataset.  To reduce the overall model parameters and speed up the model convergence, all components within the abovementioned parallel modules share the same parameters.

During training, all models are trained with the Adam optimizer ($\beta_1=0.9$, $\beta_2=0.99$). The learning rate is warmed up over the first 20 epochs to the largest value of $4\times 10^{-4}$, and linearly decreases to $4 \times 10^{-5}$ at the $100^{th}$ epoch. The batch size is set to 4. All models are trained on two NVIDIA GeForce RTX 3090 GPUs with 24 GB memory and inferred on one device.

The extracted expert features, excluding Chord Types, are continuous values, hence we first quantize them into discrete tokens following the quantization operations of FIGARO model. The vocabulary size and embedding size of individual expert features are listed in Table~\ref{feature_vocab}. To extract the learned features defined in Eq.~\eqref{learned_feature}, we adopt the similar architecture and hyperparameters as used in the original FIGARO for training a separate VQ-VAE model. To create a more lightweight version, we reduce the model dimension, latent group size, and codebook size to 256, 8, and 1024, respectively.

\begin{table}[!t]
\scriptsize
\caption{The vocabulary size and embedding size for expert features.}
\label{feature_vocab}
\centering
\begin{tabular}{lcc}
\toprule
\textbf{Type} & \textbf{Vocabulary Size} & \textbf{Embedding Size} \\
\midrule
Chord Types & 133 & 256 \\
Drum Types & 32 & 64 \\
Drum Density & 50 & 128 \\
Note Density & 66 & 128 \\
Mean Pitch & 34 & 64 \\
Mean Duration & 30 & 64 \\
Mean Velocity & 34 & 64 \\
\bottomrule
\end{tabular}
\end{table}

\section{Experimental Results and Analysis}
\label{5_results}
\subsection{Objective Evaluation}
Here, we quantitatively evaluate the performance of our proposed BandCondiNet through a detailed model comparison as well as several ablation studies. For each configuration examined, we first select a reference excerpt consisting of $B$ bars from the testing dataset, and we extract the predefined features as input conditions. Then, we feed conditional models with these features to generate $B$-bars of music. For unconditional generation models, we feed the model only a seeding token of [BOS] to generate the new sample containing $B$ bars. We refer to these generated samples as the \emph{cover} of the original reference music. We randomly select 500 reference music pieces from testing dataset, and the average values of objective metrics calculated between the 500 pairs of reference music pieces and their \emph{cover} are used to determine the final performance scores. We apply the following two categories of objective metrics, namely \emph{fidelity} and \emph{inference speed}, to evaluate the generated music.

For \emph{fidelity}, we adopt eight metrics from previous studies \citep{von2023figaro, wu2023musemorphose}, which are calculated at the bar-level or beat-level. A concise overview of these metrics is presented as below.
\begin{itemize}
    \item Note Density Error (NDE): the normalized root mean square error of bar-level note density. 
    \item Overlapping Area of Pitch (OAP), Duration (OAD), and Velocity (OAV): a similarity metric of musical elements that computes, on a bar-wise level, the overlap between the distributions of the pitch, duration, and velocity respectively in the \emph{cover} and reference. Drum notes are excluded from the computation of OAP and OAV.
    \item Cosine Similarity of Chroma (CCS) and Grooving (GCS): bar-wise cosine similarity of chroma \citep{takuya1999realtime} and grooving vectors \citep{dixon2004towards}, quantifying the closeness of two objectives in tone and rhythm, respectively. Here, the chroma vector represents the number of onsets for each of the 12 pitch classes within an octave, and the grooving vector represents the number of note onsets that occur at each of the 48 sub-beats.
    \item Chord Accuracy (CA): measures if the bar-level chords of the reference match the \emph{cover}'s chords sequence, reflecting the performance of inter-track harmony to some degree.
    \item Self-Similarity Matrix Distance (SSMD): reflects the similarity of the two music's overall music structure through beat-level Constant-Q chromagram features.
\end{itemize}

For \emph{inference speed}, we record how much time it takes for the model to generate each token (Tok/Sec) and each note (Note/Sec). This allows us to evaluate the computational efficiency when generating music samples.

\subsubsection{Comparison with Benchmarks}
\begin{table}[!t]
\scriptsize
\caption{Results of the \emph{fidelity}-related metrics for model comparison. $\downarrow$ means the lower the value, the better the result; $\uparrow$ means the larger the value, the better the result. The best values are highlighted in \textbf{bold}, and the second best is \underline{underlined}.}
\label{obj_eva_1}
\centering
\begin{tabular}{clcccccccc}
\toprule
\textbf{Datasets} & \textbf{Models} & \textbf{NDE} $\downarrow$ & \textbf{OAP} $\uparrow$ & \textbf{OAD} $\uparrow$ & \textbf{OAV} $\uparrow$ & \textbf{CCS} $\uparrow$ & \textbf{GCS} $\uparrow$ & \textbf{CA} $\uparrow$ & \textbf{SSMD} $\downarrow$ \\ \midrule
\multirow{4}{*}{32-bar}  & Music Transformer & 0.4741 & 0.7114 & 0.6222 & 0.6942 & 0.4479 & 0.8034 & 0.0821 & 0.1905 \\ 
                         & BandCondiNet-base & 0.4452 & 0.7577 & 0.6690 & 0.7218 & 0.4926 & 0.8207 & 0.0755 & 0.1805 \\ 
                         & MuseMorphose      & 0.3293 & 0.7852 & 0.7087 & 0.7646 & 0.5938 & 0.8164 & 0.2101 & 0.1597 \\ 
                         & FIGARO            & \underline{0.0944} & \underline{0.8753} & \underline{0.8574} & \underline{0.8733} & \underline{0.8002} & \underline{0.8359} & \textbf{0.5119} & \underline{0.1525} \\ 
                         & BandCondiNet (Ours) & \textbf{0.0817}& \textbf{0.8973} & \textbf{0.8776} & \textbf{0.9010} & \textbf{0.8537} & \textbf{0.8930} & \underline{0.4946} & \textbf{0.1177} \\ \midrule
\multirow{4}{*}{64-bar}  & Music Transformer & 0.5798 & 0.6982 & 0.6155 & 0.6766 & 0.4492 & 0.7867 & 0.0845 & 0.1848 \\ 
                         & BandCondiNet-base & 0.4719 & 0.7525 & 0.6651 & 0.7219 & 0.4996 & \underline{0.8314} & 0.0732 & 0.1709 \\ 
                         & MuseMorphose      & 0.3715 & 0.7819 & 0.6882 & 0.7583 & 0.5427 & 0.8211 & 0.1377 & 0.1716 \\ 
                         & FIGARO            & \underline{0.2243} & \underline{0.8344} & \underline{0.7797} & \underline{0.8157} & \underline{0.6857} & 0.8284 & \underline{0.2941} & \underline{0.1637} \\ 
                         & BandCondiNet (Ours) & \textbf{0.1441}& \textbf{0.8507} & \textbf{0.8247} & \textbf{0.8557} & \textbf{0.7978} & \textbf{0.8914} & \textbf{0.4024} & \textbf{0.1203}  \\ 
\bottomrule
\end{tabular}
\end{table}

\begin{table}[!t]
\scriptsize
\caption{Results of the \emph{inference speed}-related metrics for model comparison. All markings in this table have the same meanings as the Table~\ref{obj_eva_1}.}
\label{obj_eva_2}
\centering
\begin{tabular}{clcc}
\toprule
\textbf{Datasets} & \textbf{Models} & \textbf{Tok/Sec} $\uparrow$ & \textbf{Note/Sec} $\uparrow$ \\ \midrule
\multirow{4}{*}{32-bar} & Music Transformer & 161.08 & 120.06 \\ 
                        & BandCondiNet-base & \textbf{941.74} & \textbf{310.59} \\ 
                        & MuseMorphose      & 106.29 & 75.64 \\ 
                        & FIGARO            & 119.85 & 84.87 \\ 
                        & BandCondiNet (Ours) & \underline{467.38} & \underline{156.09} \\ \midrule
\multirow{4}{*}{64-bar} & Music Transformer & 74.64  & 51.58   \\ 
                        & BandCondiNet-base & \textbf{726.42} & \textbf{242.52}  \\ 
                        & MuseMorphose      & 53.11  & 38.73  \\ 
                        & FIGARO            & 56.59  & 39.25  \\ 
                        & BandCondiNet (Ours) & \underline{381.80} & \underline{126.63}  \\ \bottomrule
\end{tabular}
\end{table}

We first evaluate the performance of our proposed BandCondiNet against benchmark models. The detailed introduction of all benchmark models could be found in Section~\ref{4_3_benchmark_models}. For \emph{fidelity}-related metrics, Table~\ref{obj_eva_1} demonstrates that BandCondiNet outperforms other models on most metrics across both datasets. One exception is its slightly lower CA score on the 32-bar dataset, although it achieves the highest score on the longer dataset.

In terms of the results of \emph{inference speed} shown in Table~\ref{obj_eva_2}, models with parallel Transformer architectures achieve better performance, due to their capacity to process multiple sequences simultaneously instead of directly modeling a lengthy sequence. In particular, the BandCondiNet-base model achieves the best speed by a huge margin, which can be attributed to the smaller number of parameters and an entirely parallel decoding strategy. In addition, our BandCondiNet also shows great performance in terms of \emph{inference speed} among the conditional models. 

Moreover, when comparing the performance of the same model on our two datasets of different length sequences, we find that 1) all models experience a reduction in performance on the 64-bar dataset, revealing the challenge of generating longer music; 2) BandCondiNet exhibits the least performance degradation on the longer dataset, particularly in terms of CA and SSMD, demonstrating its robustness and potential of long sequence modeling.

\subsubsection{Ablation Study of BandCondiNet Components}
\begin{table*}[!t]
\scriptsize
\caption{Results of the \emph{fidelity}-related metrics for ablation study of BandCondiNet components. $\downarrow$ means the lower the value, the better the result; $\uparrow$ means the larger the value, the better the result. The best values are highlighted in \textbf{bold}, and the second best is \underline{underlined}; The \emph{italic} value indicates that the best score doesn't achieve significant improvements over that value (i.e., $p\geq 0.05$ with the one-tailed $t$-test). In contrast, the absence of the \emph{italic} font means the best score is significantly superior to other values.}
\label{ablation_study_components_1}
\centering
\begin{tabular}{clcccccccccc}
\toprule
\textbf{Datasets} & \textbf{Models} & \textbf{NDE} $\downarrow$ & \textbf{OAP} $\uparrow$ & \textbf{OAD} $\uparrow$ & \textbf{OAV} $\uparrow$ & \textbf{CCS} $\uparrow$ & \textbf{GCS $\uparrow$} & \textbf{CA} $\uparrow$ & \textbf{SSMD} $\downarrow$ \\ \midrule
\multirow{4}{*}{32-bar} & BandCondiNet      & \textbf{0.0817} & \multicolumn{1}{l}{\underline{\emph{0.8973}}} & \multicolumn{1}{l}{\textbf{0.8776}} &       \multicolumn{1}{l}{\emph{0.9010}} & \multicolumn{1}{l}{\underline{\emph{0.8537}}} & \multicolumn{1}{l}{\textbf{0.8930}} & \multicolumn{1}{l}{\textbf{0.4946}} & \multicolumn{1}{l}{\textbf{0.1177}} \\ 
                        & \quad w/o SEA   & 0.0935 & \multicolumn{1}{l}{\textbf{0.8992}} & \multicolumn{1}{l}{0.8657} & \multicolumn{1}{l}{\textbf{0.9034}} & \multicolumn{1}{l}{\textbf{0.8543}} & \multicolumn{1}{l}{\underline{\emph{0.8903}}} & \multicolumn{1}{l}{\underline{\emph{0.4906}}} & \multicolumn{1}{l}{0.1288} \\  
                        & \quad w/o CTT     & \underline{0.0922} & \multicolumn{1}{l}{\emph{0.8964}} & \multicolumn{1}{l}{\underline{\emph{0.8709}}} & \multicolumn{1}{l}{\underline{\emph{0.9015}}} & \multicolumn{1}{l}{\emph{0.8519}} & \multicolumn{1}{l}{0.8821} & \multicolumn{1}{l}{0.4844} & \multicolumn{1}{l}{\underline{\emph{0.1210}}}  \\  
                        & \quad w/o ALL & 0.1021 & \multicolumn{1}{l}{0.8862} & \multicolumn{1}{l}{0.8571} & \multicolumn{1}{l}{0.8833} & \multicolumn{1}{l}{0.8378} & \multicolumn{1}{l}{\emph{0.8874}} & \multicolumn{1}{l}{0.4715} & \multicolumn{1}{l}{0.1409}  \\ 
                        \midrule 
\multirow{4}{*}{64-bar} & BandCondiNet      & \textbf{0.1441} & \multicolumn{1}{l}{\textbf{0.8507}} & \multicolumn{1}{l}{\textbf{0.8247}} & \multicolumn{1}{l}{\textbf{0.8557}} & \multicolumn{1}{l}{\underline{\emph{0.7978}}} & \multicolumn{1}{l}{\textbf{0.8914}} & \multicolumn{1}{l}{\textbf{0.4024}} & \multicolumn{1}{l}{\textbf{0.1203}} \\ 
                        & \quad w/o SEA         & 0.1684 & \multicolumn{1}{l}{\underline{0.8453}} & \multicolumn{1}{l}{0.8154} & \multicolumn{1}{l}{0.8407} & \multicolumn{1}{l}{\textbf{0.7981}} & \multicolumn{1}{l}{\emph{0.8897}} & \multicolumn{1}{l}{\underline{\emph{0.4021}}} & \multicolumn{1}{l}{\underline{0.1321}} \\ 
                        & \quad w/o CTT           & \underline{0.1574} & \multicolumn{1}{l}{0.8392} & \multicolumn{1}{l}{\underline{0.8178}} & \multicolumn{1}{l}{\underline{0.8452}} & \multicolumn{1}{l}{0.7897} & \multicolumn{1}{l}{\underline{\emph{0.8902}}} & \multicolumn{1}{l}{0.3855} & \multicolumn{1}{l}{0.1319} \\ 
                        & \quad w/o ALL & 0.1702 & \multicolumn{1}{l}{0.8311} & \multicolumn{1}{l}{0.8069} & \multicolumn{1}{l}{0.8346} & \multicolumn{1}{l}{0.7620} & \multicolumn{1}{l}{0.8825} & \multicolumn{1}{l}{0.3612} & \multicolumn{1}{l}{0.1581} \\ 
                        \bottomrule
\end{tabular}
\end{table*}

\begin{table}[!t]
\scriptsize
\caption{Results of the \emph{inference speed}-related metrics for ablation study of BandCondiNet components. All markings in this table have the same meanings as the Table~\ref{ablation_study_components_1}.}
\label{ablation_study_components_2}
\centering
\begin{tabular}{clcc}
\toprule
\textbf{Datasets} & \textbf{Models} & \textbf{Tok/Sec} $\uparrow$ & \textbf{Note/Sec} $\uparrow$ \\ \midrule
\multirow{4}{*}{32-bar} & BandCondiNet           & 467.38 & 156.09 \\ 
                        & \quad w/o SEA          & 480.54 & 159.78 \\ 
                        & \quad w/o CTT          & \underline{513.59} & \underline{170.15} \\ 
                        & \quad w/o ALL          & \textbf{541.46} & \textbf{178.34} \\ \midrule
\multirow{4}{*}{64-bar} & BandCondiNet           & 381.80  & 126.63   \\ 
                        & \quad w/o SEA          & 412.47 & 136.78  \\ 
                        & \quad w/o CTT          & \underline{417.51}  & \underline{138.43}  \\ 
                        & \quad w/o ALL          & \textbf{449.81} & \textbf{148.74}  \\ \bottomrule
\end{tabular}
\end{table}

We analyze the effectiveness of the various BandCondiNet components. All ablation studies are conducted on the 32-bar and 64-bar datasets with the following three configurations: 1) \textbf{w/o SEA}, which only uses vanilla self-attention instead of the proposed Structure Enhanced Attention in the Bottom Decoders and Top Decoders; 2) \textbf{w/o CTT}, which only deletes the Cross-Track Transformer in the decoder block; 3) \textbf{w/o ALL}, which uses vanilla self-attention modules and drops the CTT module.

Tables~\ref{ablation_study_components_1} and~\ref{ablation_study_components_2} show that: 1) BandCondiNet with complete architecture performs best on most \emph{fidelity}-related metrics, especially on the longer music dataset; 2) Removing the SEA module results in worse scores of most \emph{fidelity}-related metrics (in particular, the SSMD scores degrade significantly in both two datasets), indicating that our proposed SEA could enhance the music structure modeling; 3) Removing CTT modules decreases the values of most \emph{fidelity}-related metrics (CA scores degrade significantly in both two datasets especially) but increases the inference speed. This may be attributed to the fact that CTT facilitates the interaction between tracks but introduces extra computation.

\subsubsection{Ablation Study of Input Conditions}
Finally, we evaluate the effectiveness of the input conditions. Here, we mainly focus on the comparison between using the bar-level conditions designed by FIGARO and our proposed multi-view conditions, since the FIGARO and BandcondiNet significantly outperform other models in Table~\ref{obj_eva_1} and~\ref{obj_eva_2}. We apply both bar-level conditions and our multi-view conditions to both FIGARO and BandCondiNet. Additionally, we also examine the impact of using partial multi-view features (i.e., expert features only or learned features only) with BandCondiNet. This ablation study focuses solely on the \emph{fidelity}-related metrics, as \emph{inference speed} would behave similarly under the same model architecture.

Table~\ref{ablation_study_features} shows the following findings: 
1) Multi-view conditions consistently outperform bar-level conditions, especially on the 64-bar dataset, across both FIGARO and BandCondiNet. 
This indicates that our multi-view features enhance the generation of multitrack music with high-fidelity; 
2) The overall performance of BandCondiNet decreases when either expert features or learned features are removed, though is still surpasses the case when using bar-level conditions.

A significant performance decline occurs when bar-level conditions are applied to BandCondiNet compared to using multi-view conditions, while FIGARO experiences only a slight decrease in performance under the same conditions. This discrepancy likely arises from the fact that bar-level conditions are averages features across all tracks, including information from other instruments. When BandCondiNet receives these bar-level conditions, its parallel decoders are fed with identical signals, introducing noise (i.e., information from other tracks) during the generation for a specific instrument. As a result, the bar-level conditions not only fail to improve BandCondiNet's performance but actually impair it. In contrast, FIGARO is designed to generate all tracks collectively, the bar-level conditions would not be regarded as noise and would be fully leveraged.

\begin{table*}[!t]
\tiny
\caption{Ablation study of input conditions. All markings in this table have the same meanings as the Table~\ref{ablation_study_components_1}.}
\label{ablation_study_features}
\centering
\begin{tabular}{clcccccccccc}
\toprule
\textbf{Datasets} & \textbf{Configurations} & \textbf{NDE} $\downarrow$ & \textbf{OAP} $\uparrow$ & \textbf{OAD} $\uparrow$ & \textbf{OAV} $\uparrow$ & \textbf{CCS} $\uparrow$ & \textbf{GCS $\uparrow$} & \textbf{CA} $\uparrow$ & \textbf{SSMD} $\downarrow$ \\ \midrule
\multirow{6}{*}{32-bar} & [Bar-level, FIGARO]   & \underline{0.0944} & \multicolumn{1}{l}{0.8753} & \multicolumn{1}{l}{0.8574} & 0.8733 & 0.8002 & \multicolumn{1}{l}{0.8359} & \multicolumn{1}{l}{\underline{\emph{0.5119}}} & 0.1525 \\ 
                        & [Multi-view, FIGARO] & 0.1079 & \multicolumn{1}{l}{\underline{0.8821}} & \multicolumn{1}{l}{\underline{0.8608}} & \underline{0.8944} & \underline{0.8156} & \multicolumn{1}{l}{\underline{\emph{0.8884}}} & \multicolumn{1}{l}{\textbf{0.5156}} & 0.1444 \\ \cmidrule(lr){2-10} 
                        & [Bar-level, BandCondiNet]    & 0.2381 & \multicolumn{1}{l}{0.8320} & \multicolumn{1}{l}{0.7597} & 0.7992 & 0.7727 & \multicolumn{1}{l}{0.8343} & \multicolumn{1}{l}{0.4432} & 0.1489  \\ 
                        & [Multi-view(w/o $F^{exp}$), BandCondiNet] & 0.1128 & \multicolumn{1}{l}{0.8701} & \multicolumn{1}{l}{0.7997} & 0.8726 & 0.7545 & \multicolumn{1}{l}{0.8749} & \multicolumn{1}{l}{0.2891} & \underline{0.1312} \\
                        & [Multi-view(w/o $F^{vq}$), BandCondiNet] & 0.1105 & \multicolumn{1}{l}{0.8343} & \multicolumn{1}{l}{0.8499} & 0.8798 & 0.7537 & \multicolumn{1}{l}{0.8393} & \multicolumn{1}{l}{0.4537} & 0.1402 \\
                        & [Multi-view, BandCondiNet]    & \textbf{0.0817} & \multicolumn{1}{l}{\textbf{0.8973}} & \multicolumn{1}{l}{\textbf{0.8776}} & \textbf{0.9010} & \textbf{0.8537} & \multicolumn{1}{l}{\textbf{0.8930}} & \multicolumn{1}{l}{0.4946} & \textbf{0.1177} \\ 
                        \midrule 
\multirow{6}{*}{64-bar} & [Bar-level, FIGARO]  & 0.2243 & \multicolumn{1}{l}{0.8344} & \multicolumn{1}{l}{0.7797} & 0.8157 & 0.6857 & \multicolumn{1}{l}{0.8284} & \multicolumn{1}{l}{0.2941} & 0.1637   \\ 
                        & [Multi-view, FIGARO]  & 0.2091 & \multicolumn{1}{l}{\underline{0.8379}} & \multicolumn{1}{l}{0.7824} & \underline{0.8313} & \underline{0.7219} & \multicolumn{1}{l}{\underline{0.8847}} & \multicolumn{1}{l}{0.3461} & 0.1496  \\ \cmidrule(lr){2-10}
                        & [Bar-level, BandCondiNet] & 0.3653 & \multicolumn{1}{l}{0.7764} & \multicolumn{1}{l}{0.6623} & 0.7314 & 0.4978 & \multicolumn{1}{l}{0.8373} & \multicolumn{1}{l}{0.2168} & 0.1652  \\
                        & [Multi-view(w/o $F^{exp}$), BandCondiNet] & \underline{0.1753} & \multicolumn{1}{l}{0.8250} & \multicolumn{1}{l}{0.7511} & 0.8229 & 0.7190 & \multicolumn{1}{l}{0.8764} & \multicolumn{1}{l}{0.2438} & \underline{0.1427} \\
                        & [Multi-view(w/o $F^{vq}$), BandCondiNet] & 0.1947 & \multicolumn{1}{l}{0.7922} & \multicolumn{1}{l}{\underline{0.7871}} & 0.8229 & 0.6974 & \multicolumn{1}{l}{0.8322} & \multicolumn{1}{l}{\underline{0.3517}} & 0.1491 \\
                        & [Multi-view, BandCondiNet] & \textbf{0.1441} & \multicolumn{1}{l}{\textbf{0.8507}} & \multicolumn{1}{l}{\textbf{0.8247}} & \textbf{0.8557} & \textbf{0.7978} & \multicolumn{1}{l}{\textbf{0.8914}} & \multicolumn{1}{l}{\textbf{0.4024}} & \textbf{0.1203}  \\
                        \bottomrule
\end{tabular}
\end{table*}

\subsection{Subjective Evaluation}

\begin{figure}[!t]
    \centering
    \subfigure[Results for the 32-bar dataset.\label{sub_eva_32}]{
    \includegraphics[width=0.8\textwidth]{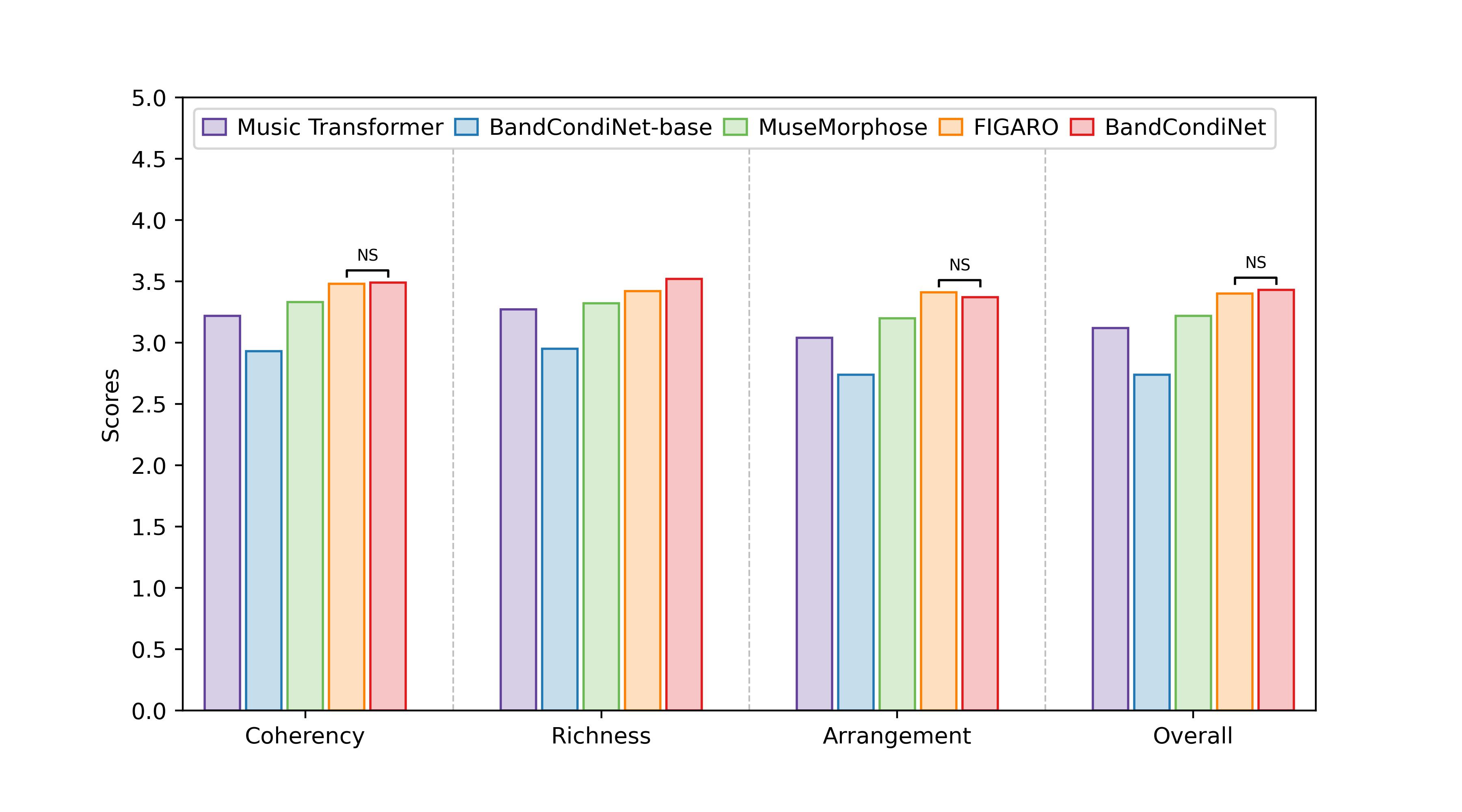}
    }
    \subfigure[Results for the 64-bar dataset.\label{sub_eva_64}]{
    \includegraphics[width=0.8\textwidth]{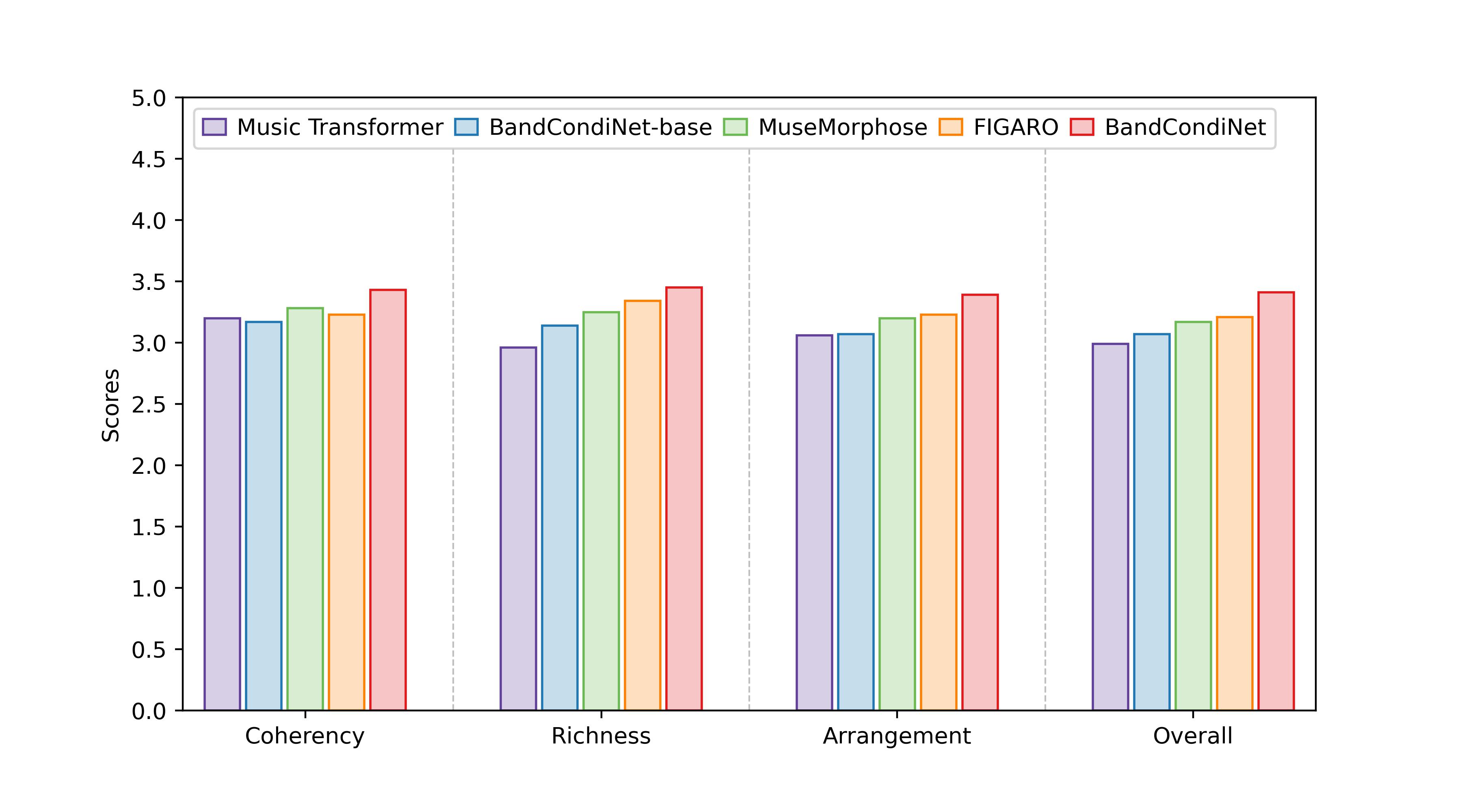}
    }
    \caption{Subjective evaluation for samples generated by BandCondiNet and the four benchmark models. The mark of NS between the two bars represents that there is no significant difference ($p\geq 0.05$ with the one-tailed $t$-test) between the two values.}
	\label{sub_eva}
\end{figure}

In addition to objective evaluation, we also conducted two listening tests to assess the generated samples' quality for models trained on both of our datasets. Each survey contains six sets of generated samples. Each set contains one piece of reference music and five \emph{cover} samples generated by BandCondiNet and the four benchmark models. These samples are randomly ordered. We synthesize all MIDIs using the default sound font in the Overtune 5 software\footnote{\url{https://sonicscores.com/overture/}}. In the questionnaire, each participant is required to rate each \emph{cover} sample based on a 5-point Likert scale from 1 (very low) to 5 (very high) in terms of the following criteria: 
1) \textbf{Coherency}: How well does the music sound temporally coherent and fluent? 
2) \textbf{Richness}: How well does the music demonstrate diversity, particularly in its melodic development, texture, and structure?
3) \textbf{Arrangement}: Do the notes from different instruments have proper coordination? Do all the voice relationships sound musically reasonable?
4) \textbf{Overall}: How good is the overall musical quality?

We recruited a total of 73 participants (20 female and 53 males, aged from 20 to 50) from a multidisciplinary university to ensure a diverse listener pool. Among these, 29 participants (approximately 40\%) had more than three years of musical experience. This group notably included four music professors from the university's music education department, whose expert opinions ensured a rigorous technical evaluation. The remaining 44 participants were casual music enthusiasts without formal theory training, providing essential feedback on the music's general listenability and appeal from a non-expert, general audience perspective.
Fig.~\ref{sub_eva} shows the results of the two subjective evaluations. It can be seen that our BandCondiNet model achieves the best scores for Richness when trained on the 32-bar dataset, and reaches similar performance to the state-of-the-art FIGARO model in the remaining metrics. Looking at the results are even better when the models are trained on the 64-bar dataset. In this case, the samples generated by BandCondiNet obtain significantly higher scores than samples generated by the benchmark models, and this across all criteria. This confirms that our model excels in generating long sequences without losing much fidelity.

\subsection{Visual Analysis}
As a supplemental analysis, we visually compare a sample of the \emph{cover} music generated by BandCondiNet and the FIGARO benchmark with the reference music from the testing dataset. Here, we omit unconditional models and MuseMorphose as they did not show competitive performance on the objective and subjective tests.

\begin{figure}[!t]
    \centering
    \subfigure[The reference.\label{case_ori}]{
    \includegraphics[width=0.45\textwidth]{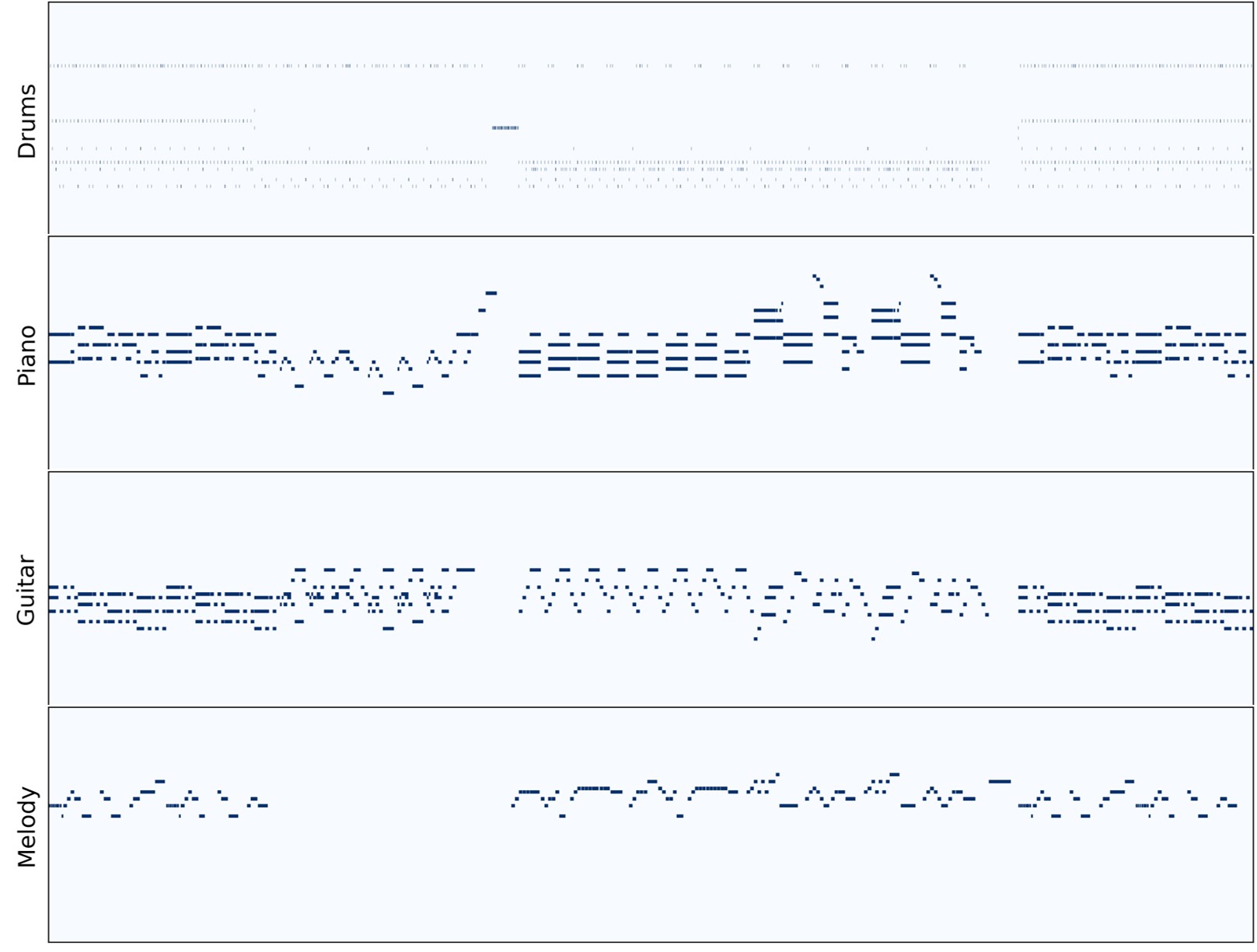}
    }
    \subfigure[The \emph{cover} by FIGARO model.\label{case_figaro}]{
    \includegraphics[width=0.45\textwidth]{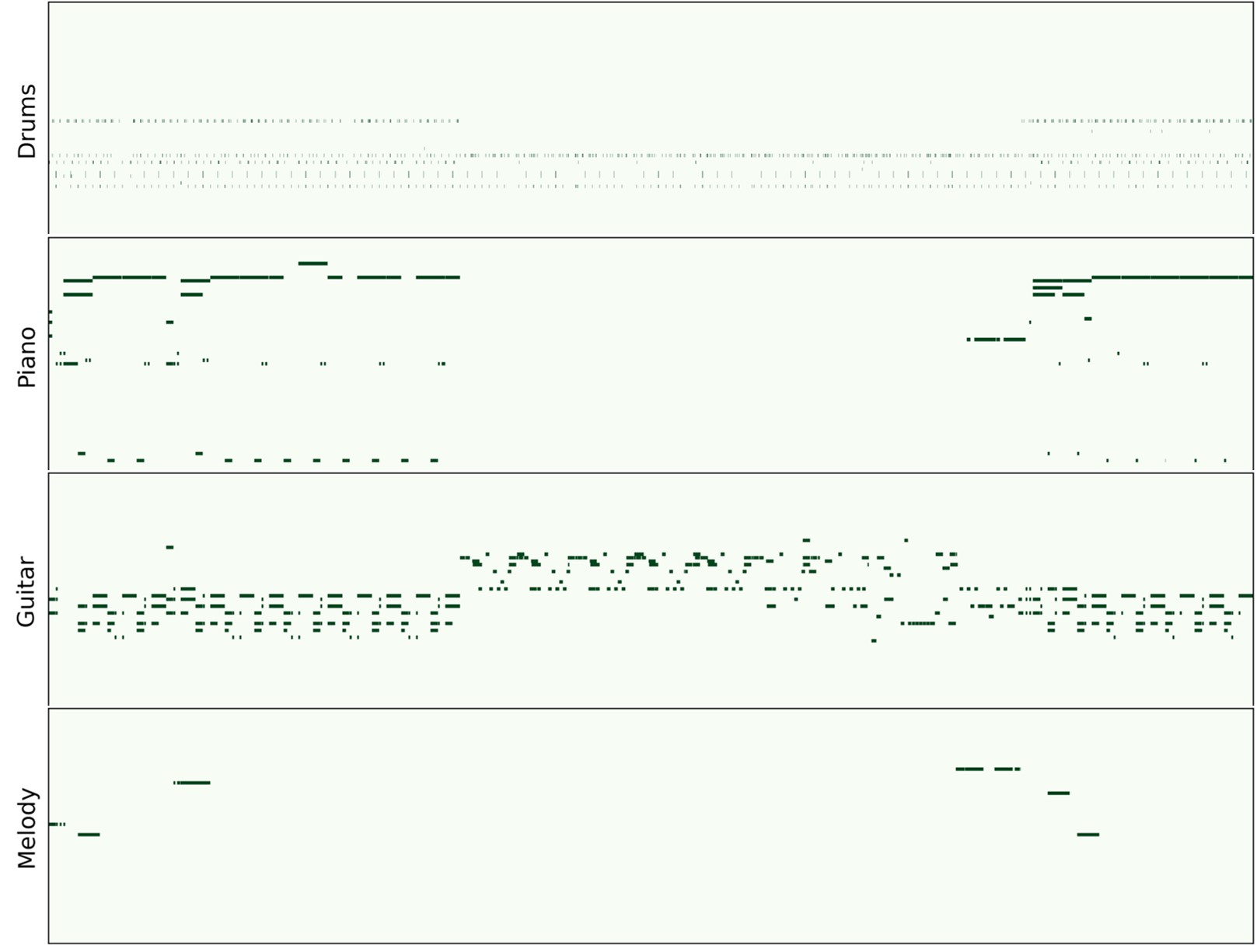}
    }
    \subfigure[The \emph{cover} generated by BandCondiNet.\label{case_bandcondinet}]{
    \includegraphics[width=0.45\textwidth]{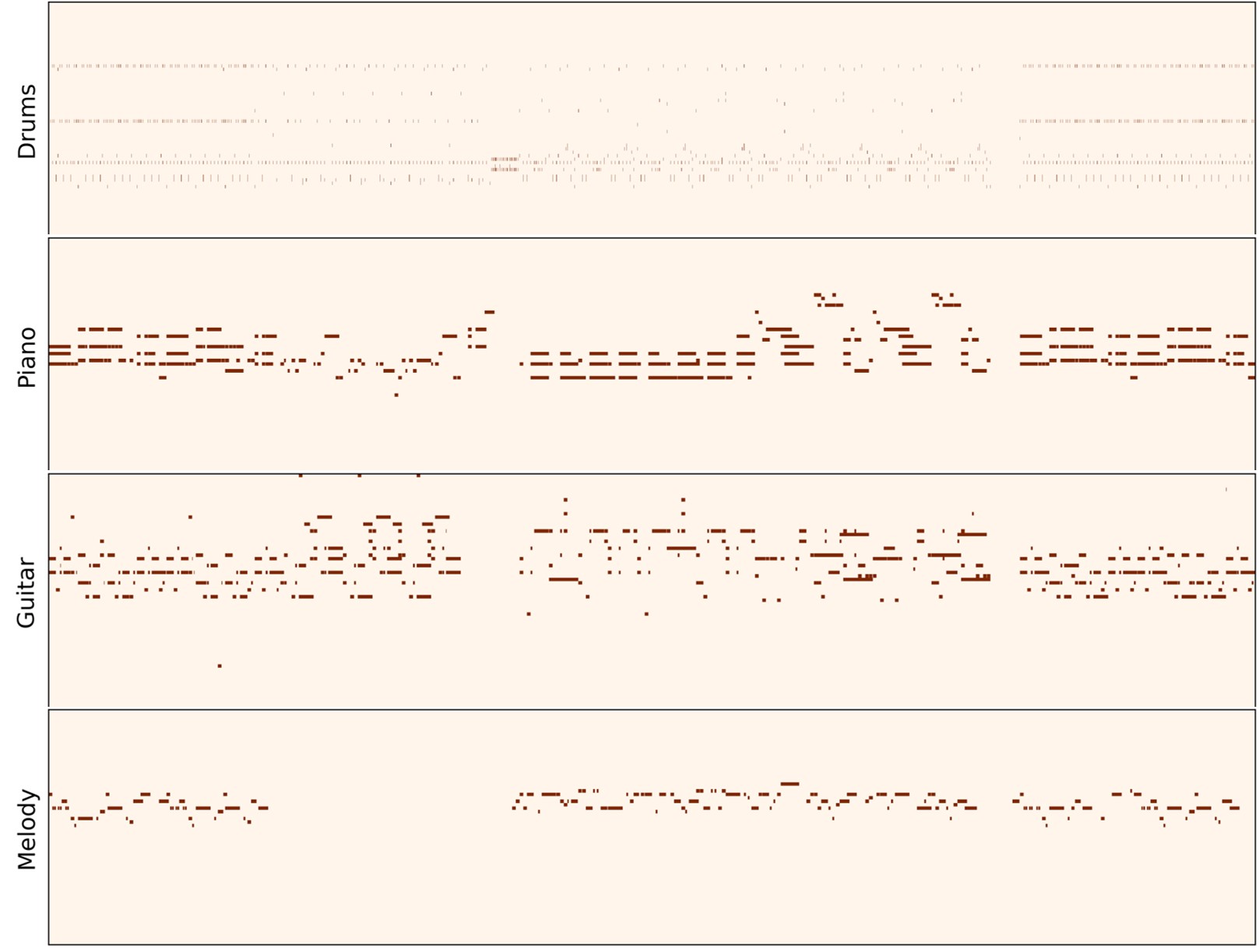}
    }
    \caption{Visualized pianoroll of reference music, \emph{cover} generated by FIGARO, and \emph{cover} generated by BandCondiNet. The x-axis represents bar numbers while the y-axis represents pitch range.}
	\label{case_study}
\end{figure}

Fig.~\ref{case_study} shows three sets of visualized pianoroll corresponding to the reference, the \emph{cover} generated by FIGARO and the \emph{cover} generated by BandCondiNet. The reference music is picked from the 64-bar dataset. It has 41 bars in total and 5 phrases with the manually annotated structure of A8x8B8C9A8 (`A', `B', and `C' denote a phrase with melody; `x' represents a bridge). 

As we can see, the structure of the \emph{cover} of BandCondiNet is approximately in line with that of the reference. More interestingly, it also presents various repeating melodies and textures within a phrase or between phrases, even spread over long distances. These findings demonstrate that BandCondiNet is able to learn the structure of music guided by the multi-view conditions and develop interesting patterns. In contrast, the melody generated by the FIGARO model collapses and the textures sound boring since they repeat certain fixed patterns. This is mostly due to the fact that the bar-level conditions fed into FIGARO are mixtures summarized from all instruments, and contain fewer instructions for each specific track\footnote{More music samples are available at \url{https://chinglohsiu.github.io/files/bandcondinet.html}}.

\section{Conclusion and Discussion}
\label{6_conclusion}
In this paper, we introduce three-dimensional multi-view features that are extracted from both the bar-view (per bar) and track-view (per track) of a musical piece, offering fine-grained details that enhance the fidelity of generated music. We propose a parallel Transformers-based architecture called BandCondiNet for conditional music generation. This architecture fully integrates the multi-view features and facilitates a more effective modeling of the music structure and inter-track harmony, due to the proposed Structure Enhanced Attention and Cross-Track Transformer modules. Both objective and subjective evaluations show that BandCondiNet generates \emph{covers} with higher musical fidelity, superior structural coherence, enhanced inter-track harmony and faster inference speeds compared to other state-of-the-art conditional models, and also exhibits robustness in generating long music sequences. Finally, a visual case study is conducted to validate these findings.

While BandCondiNet presents prominent improvements in the fidelity and quality of the generated music, we identify three main limitations as well as essential directions for future work. First, the instrumentation in the generated \emph{cover} remains fixed to the reference (a set of six commonly used instruments in popular music), limiting flexibility in orchestration and instrumentation. This configuration of track compression simplifies multitrack modeling but sacrifices the rich timbre of popular music. Second, given the multi-view features of certain patterns such as simple bass patterns and  block chords, BandCondiNet sometimes replicates the reference instead of generating a novel version. This balance between fidelity and creativity of the conditional music generation model is a challenge for future work. Third, when considering to apply our model for controllable music generation, the expert features from the multi-view features enable some level of user control due to their human-interpretability. In contrast, the abstract learned features must be learned from an existing reference, which is less user-friendly and hard to manipulate directly.

In the future, we plan to design an advanced conditional model to tackle flexible instrumentation and encourage the generation of creative, yet stylistically faithful music. Moreover, we also intend to develop a middle-ware module, connecting the multi-view features (especially for learned features) with user-friendly input. In this way, user expectations will be first translated into meaningful conditions that guide the creative process more effectively.

\section*{CRediT authorship contribution statement}
\textbf{Jing Luo:} Conceptualization, Methodology, Validation, Writing - Original Draft. \textbf{Xinyu Yang}: Supervision, Project administration. \textbf{Dorien Herremans:} Supervision, Funding acquisition, Writing - Review \& Editing.

\section*{Data availability}
The data that support the findings of this study are available from the corresponding author upon reasonable request.

\section*{Acknowledgment}
The authors would like to thank all those who participated in the listening tests of subjective evaluation. This work is funded by the Chinese Scholarship Council no. 202006280348 and the SUTD Kickstarter Initiative no. SKI 2021\_04\_06.



\bibliographystyle{model5-names} 
\bibliography{bandcondinet-ref}\biboptions{authoryear}

\begin{thebibliography}{53}
\expandafter\ifx\csname natexlab\endcsname\relax\def\natexlab#1{#1}\fi
\providecommand{\bibinfo}[2]{#2}
\ifx\xfnm\relax \def\xfnm[#1]{\unskip,\space#1}\fi
\bibitem[{Bao \& Sun(2023)}]{bao2023generating}
\bibinfo{author}{Bao, C.}, \& \bibinfo{author}{Sun, Q.} (\bibinfo{year}{2023}).
\newblock \bibinfo{title}{Generating music with emotions}.
\newblock {\it \bibinfo{journal}{IEEE Transactions on Multimedia}\/},  {\it \bibinfo{volume}{25}\/}, \bibinfo{pages}{3602--3614}.
\bibitem[{Belkin(2018)}]{belkin2018musical}
\bibinfo{author}{Belkin, A.} (\bibinfo{year}{2018}).
\newblock {\it \bibinfo{title}{Musical composition: Craft and art}\/}.
\newblock \bibinfo{publisher}{Yale University Press}.
\bibitem[{Bertin-Mahieux et~al.(2011)Bertin-Mahieux, Ellis, Whitman \& Lamere}]{bertin2011million}
\bibinfo{author}{Bertin-Mahieux, T.}, \bibinfo{author}{Ellis, D.~P.}, \bibinfo{author}{Whitman, B.}, \& \bibinfo{author}{Lamere, P.} (\bibinfo{year}{2011}).
\newblock \bibinfo{title}{The million song dataset.}
\newblock In {\it \bibinfo{booktitle}{Ismir}\/} (p.~\bibinfo{pages}{10}).
\newblock volume~\bibinfo{volume}{2}.
\bibitem[{Choi \& Lee(2023)}]{choi2023pop2piano}
\bibinfo{author}{Choi, J.}, \& \bibinfo{author}{Lee, K.} (\bibinfo{year}{2023}).
\newblock \bibinfo{title}{Pop2piano: Pop audio-based piano cover generation}.
\newblock In {\it \bibinfo{booktitle}{ICASSP 2023-2023 IEEE International Conference on Acoustics, Speech and Signal Processing (ICASSP)}\/} (pp. \bibinfo{pages}{1--5}).
\newblock \bibinfo{organization}{IEEE}.
\bibitem[{Choi et~al.(2020)Choi, Hawthorne, Simon, Dinculescu \& Engel}]{choi2020encoding}
\bibinfo{author}{Choi, K.}, \bibinfo{author}{Hawthorne, C.}, \bibinfo{author}{Simon, I.}, \bibinfo{author}{Dinculescu, M.}, \& \bibinfo{author}{Engel, J.} (\bibinfo{year}{2020}).
\newblock \bibinfo{title}{Encoding musical style with transformer autoencoders}.
\newblock In {\it \bibinfo{booktitle}{International conference on machine learning}\/} (pp. \bibinfo{pages}{1899--1908}).
\bibitem[{Civit et~al.(2022)Civit, Civit-Masot, Cuadrado \& Escalona}]{civit2022systematic}
\bibinfo{author}{Civit, M.}, \bibinfo{author}{Civit-Masot, J.}, \bibinfo{author}{Cuadrado, F.}, \& \bibinfo{author}{Escalona, M.~J.} (\bibinfo{year}{2022}).
\newblock \bibinfo{title}{A systematic review of artificial intelligence-based music generation: Scope, applications, and future trends}.
\newblock {\it \bibinfo{journal}{Expert Systems with Applications}\/},  {\it \bibinfo{volume}{209}\/}, \bibinfo{pages}{118190}.
\bibitem[{Copet et~al.(2024)Copet, Kreuk, Gat, Remez, Kant, Synnaeve, Adi \& D{\'e}fossez}]{copet2024simple}
\bibinfo{author}{Copet, J.}, \bibinfo{author}{Kreuk, F.}, \bibinfo{author}{Gat, I.}, \bibinfo{author}{Remez, T.}, \bibinfo{author}{Kant, D.}, \bibinfo{author}{Synnaeve, G.}, \bibinfo{author}{Adi, Y.}, \& \bibinfo{author}{D{\'e}fossez, A.} (\bibinfo{year}{2024}).
\newblock \bibinfo{title}{Simple and controllable music generation}.
\newblock {\it \bibinfo{journal}{Advances in Neural Information Processing Systems}\/},  {\it \bibinfo{volume}{36}\/}.
\bibitem[{Dai et~al.(2020)Dai, Zhang \& Dannenberg}]{dai2020automatic}
\bibinfo{author}{Dai, S.}, \bibinfo{author}{Zhang, H.}, \& \bibinfo{author}{Dannenberg, R.~B.} (\bibinfo{year}{2020}).
\newblock \bibinfo{title}{Automatic analysis and influence of hierarchical structure on melody, rhythm and harmony in popular music}.
\newblock {\it \bibinfo{journal}{arXiv preprint arXiv:2010.07518}\/}, .
\bibitem[{Ding \& Cui(2023)}]{ding2023museflow}
\bibinfo{author}{Ding, F.}, \& \bibinfo{author}{Cui, Y.} (\bibinfo{year}{2023}).
\newblock \bibinfo{title}{Museflow: music accompaniment generation based on flow}.
\newblock {\it \bibinfo{journal}{Applied Intelligence}\/},  (pp. \bibinfo{pages}{1--10}).
\bibitem[{Dixon et~al.(2004)Dixon, Gouyon, Widmer et~al.}]{dixon2004towards}
\bibinfo{author}{Dixon, S.}, \bibinfo{author}{Gouyon, F.}, \bibinfo{author}{Widmer, G.} et~al. (\bibinfo{year}{2004}).
\newblock \bibinfo{title}{Towards characterisation of music via rhythmic patterns.}
\newblock In {\it \bibinfo{booktitle}{ISMIR}\/}.
\bibitem[{Donahue et~al.(2019)Donahue, Mao, Li, Cottrell \& McAuley}]{Donahue2019LakhNES}
\bibinfo{author}{Donahue, C.}, \bibinfo{author}{Mao, H.~H.}, \bibinfo{author}{Li, Y.~E.}, \bibinfo{author}{Cottrell, G.~W.}, \& \bibinfo{author}{McAuley, J.~J.} (\bibinfo{year}{2019}).
\newblock \bibinfo{title}{Lakhnes: Improving multi-instrumental music generation with cross-domain pre-training}.
\newblock In {\it \bibinfo{booktitle}{Proceedings of the 20th International Society for Music Information Retrieval Conference, {ISMIR} 2019, Delft, The Netherlands, November 4-8, 2019}\/} (pp. \bibinfo{pages}{685--692}).
\bibitem[{Dong et~al.(2023)Dong, Chen, Dubnov, McAuley \& Berg-Kirkpatrick}]{dong2023multitrack}
\bibinfo{author}{Dong, H.-W.}, \bibinfo{author}{Chen, K.}, \bibinfo{author}{Dubnov, S.}, \bibinfo{author}{McAuley, J.}, \& \bibinfo{author}{Berg-Kirkpatrick, T.} (\bibinfo{year}{2023}).
\newblock \bibinfo{title}{Multitrack music transformer}.
\newblock In {\it \bibinfo{booktitle}{ICASSP 2023-2023 IEEE International Conference on Acoustics, Speech and Signal Processing (ICASSP)}\/} (pp. \bibinfo{pages}{1--5}).
\newblock \bibinfo{organization}{IEEE}.
\bibitem[{Dong et~al.(2018)Dong, Hsiao, Yang \& Yang}]{dong2018musegan}
\bibinfo{author}{Dong, H.-W.}, \bibinfo{author}{Hsiao, W.-Y.}, \bibinfo{author}{Yang, L.-C.}, \& \bibinfo{author}{Yang, Y.-H.} (\bibinfo{year}{2018}).
\newblock \bibinfo{title}{Musegan: Multi-track sequential generative adversarial networks for symbolic music generation and accompaniment}.
\newblock In {\it \bibinfo{booktitle}{Proceedings of the AAAI Conference on Artificial Intelligence}\/} (pp. \bibinfo{pages}{34--41}).
\bibitem[{Ens \& Pasquier(2020)}]{ens2020mmm}
\bibinfo{author}{Ens, J.}, \& \bibinfo{author}{Pasquier, P.} (\bibinfo{year}{2020}).
\newblock \bibinfo{title}{Mmm: Exploring conditional multi-track music generation with the transformer}.
\newblock {\it \bibinfo{journal}{arXiv preprint arXiv:2008.06048}\/}, .
\bibitem[{Esser et~al.(2021)Esser, Rombach \& Ommer}]{esser2021taming}
\bibinfo{author}{Esser, P.}, \bibinfo{author}{Rombach, R.}, \& \bibinfo{author}{Ommer, B.} (\bibinfo{year}{2021}).
\newblock \bibinfo{title}{Taming transformers for high-resolution image synthesis}.
\newblock In {\it \bibinfo{booktitle}{Proceedings of the IEEE/CVF conference on computer vision and pattern recognition}\/} (pp. \bibinfo{pages}{12873--12883}).
\bibitem[{Fradet et~al.(2023)Fradet, Gutowski, Chhel \& Briot}]{fradet2023byte}
\bibinfo{author}{Fradet, N.}, \bibinfo{author}{Gutowski, N.}, \bibinfo{author}{Chhel, F.}, \& \bibinfo{author}{Briot, J.} (\bibinfo{year}{2023}).
\newblock \bibinfo{title}{Byte pair encoding for symbolic music}.
\newblock In {\it \bibinfo{booktitle}{Proceedings of the 2023 Conference on Empirical Methods in Natural Language Processing}\/} (pp. \bibinfo{pages}{2001--2020}).
\bibitem[{Guo et~al.(2019)Guo, Herremans \& Magnusson}]{guo2019midi}
\bibinfo{author}{Guo, R.}, \bibinfo{author}{Herremans, D.}, \& \bibinfo{author}{Magnusson, T.} (\bibinfo{year}{2019}).
\newblock \bibinfo{title}{Midi miner--a python library for tonal tension and track classification}.
\newblock {\it \bibinfo{journal}{arXiv preprint arXiv:1910.02049}\/}, .
\bibitem[{Guo et~al.(2022)Guo, Simpson, Kiefer, Magnusson \& Herremans}]{guo2022musiac}
\bibinfo{author}{Guo, R.}, \bibinfo{author}{Simpson, I.}, \bibinfo{author}{Kiefer, C.}, \bibinfo{author}{Magnusson, T.}, \& \bibinfo{author}{Herremans, D.} (\bibinfo{year}{2022}).
\newblock \bibinfo{title}{Musiac: An extensible generative framework for music infilling applications with multi-level control}.
\newblock In {\it \bibinfo{booktitle}{International Conference on Computational Intelligence in Music, Sound, Art and Design (Part of EvoStar)}\/} (pp. \bibinfo{pages}{341--356}).
\bibitem[{Herremans et~al.(2017)Herremans, Chuan \& Chew}]{herremans2017functional}
\bibinfo{author}{Herremans, D.}, \bibinfo{author}{Chuan, C.-H.}, \& \bibinfo{author}{Chew, E.} (\bibinfo{year}{2017}).
\newblock \bibinfo{title}{A functional taxonomy of music generation systems}.
\newblock {\it \bibinfo{journal}{ACM Computing Surveys (CSUR)}\/},  {\it \bibinfo{volume}{50}\/}, \bibinfo{pages}{1--30}.
\bibitem[{Huang et~al.(2019)Huang, Vaswani, Uszkoreit, Simon, Hawthorne, Shazeer, Dai, Hoffman, Dinculescu \& Eck}]{Huang2019musictransformer}
\bibinfo{author}{Huang, C.~A.}, \bibinfo{author}{Vaswani, A.}, \bibinfo{author}{Uszkoreit, J.}, \bibinfo{author}{Simon, I.}, \bibinfo{author}{Hawthorne, C.}, \bibinfo{author}{Shazeer, N.}, \bibinfo{author}{Dai, A.~M.}, \bibinfo{author}{Hoffman, M.~D.}, \bibinfo{author}{Dinculescu, M.}, \& \bibinfo{author}{Eck, D.} (\bibinfo{year}{2019}).
\newblock \bibinfo{title}{Music transformer: Generating music with long-term structure}.
\newblock In {\it \bibinfo{booktitle}{7th International Conference on Learning Representations, {ICLR} 2019, New Orleans, LA, USA, May 6-9, 2019}\/} (pp. \bibinfo{pages}{1--15}).
\bibitem[{Huang et~al.(2024)Huang, Ghatare, Liu, Hu, Zhang, Sastry, Gururani, Oore \& Yue}]{huang2024symbolic}
\bibinfo{author}{Huang, Y.}, \bibinfo{author}{Ghatare, A.}, \bibinfo{author}{Liu, Y.}, \bibinfo{author}{Hu, Z.}, \bibinfo{author}{Zhang, Q.}, \bibinfo{author}{Sastry, C.~S.}, \bibinfo{author}{Gururani, S.}, \bibinfo{author}{Oore, S.}, \& \bibinfo{author}{Yue, Y.} (\bibinfo{year}{2024}).
\newblock \bibinfo{title}{Symbolic music generation with non-differentiable rule guided diffusion}.
\newblock In {\it \bibinfo{booktitle}{Forty-first International Conference on Machine Learning}\/} (pp. \bibinfo{pages}{1--26}).
\bibitem[{Huang \& Yang(2020)}]{huang2020pop}
\bibinfo{author}{Huang, Y.-S.}, \& \bibinfo{author}{Yang, Y.-H.} (\bibinfo{year}{2020}).
\newblock \bibinfo{title}{Pop music transformer: Beat-based modeling and generation of expressive pop piano compositions}.
\newblock In {\it \bibinfo{booktitle}{Proceedings of the 28th ACM international conference on multimedia}\/} (pp. \bibinfo{pages}{1180--1188}).
\bibitem[{Ji et~al.(2024)Ji, Yang \& Luo}]{2020comprehensive}
\bibinfo{author}{Ji, S.}, \bibinfo{author}{Yang, X.}, \& \bibinfo{author}{Luo, J.} (\bibinfo{year}{2024}).
\newblock \bibinfo{title}{A survey on deep learning for symbolic music generation: Representations, algorithms, evaluations, and challenges}.
\newblock {\it \bibinfo{journal}{ACM Computing Surveys}\/},  {\it \bibinfo{volume}{56}\/}, \bibinfo{pages}{7:1--7:39}.
\bibitem[{Jin et~al.(2022)Jin, Wang, Li, Tie, Tie, Liu, Yan, Li, Wang \& Huang}]{jin2022transformer}
\bibinfo{author}{Jin, C.}, \bibinfo{author}{Wang, T.}, \bibinfo{author}{Li, X.}, \bibinfo{author}{Tie, C. J.~J.}, \bibinfo{author}{Tie, Y.}, \bibinfo{author}{Liu, S.}, \bibinfo{author}{Yan, M.}, \bibinfo{author}{Li, Y.}, \bibinfo{author}{Wang, J.}, \& \bibinfo{author}{Huang, S.} (\bibinfo{year}{2022}).
\newblock \bibinfo{title}{A transformer generative adversarial network for multi-track music generation}.
\newblock {\it \bibinfo{journal}{CAAI Transactions on Intelligence Technology}\/},  {\it \bibinfo{volume}{7}\/}, \bibinfo{pages}{369--380}.
\bibitem[{Kaiser et~al.(2018)Kaiser, Bengio, Roy, Vaswani, Parmar, Uszkoreit \& Shazeer}]{kaiser2018fast}
\bibinfo{author}{Kaiser, L.}, \bibinfo{author}{Bengio, S.}, \bibinfo{author}{Roy, A.}, \bibinfo{author}{Vaswani, A.}, \bibinfo{author}{Parmar, N.}, \bibinfo{author}{Uszkoreit, J.}, \& \bibinfo{author}{Shazeer, N.} (\bibinfo{year}{2018}).
\newblock \bibinfo{title}{Fast decoding in sequence models using discrete latent variables}.
\newblock In {\it \bibinfo{booktitle}{International Conference on Machine Learning}\/} (pp. \bibinfo{pages}{2390--2399}).
\bibitem[{Kang et~al.(2024)Kang, Poria \& Herremans}]{kang2024video2music}
\bibinfo{author}{Kang, J.}, \bibinfo{author}{Poria, S.}, \& \bibinfo{author}{Herremans, D.} (\bibinfo{year}{2024}).
\newblock \bibinfo{title}{Video2music: Suitable music generation from videos using an affective multimodal transformer model}.
\newblock {\it \bibinfo{journal}{Expert Systems with Applications}\/},  {\it \bibinfo{volume}{249}\/}, \bibinfo{pages}{123640}.
\bibitem[{Le et~al.(2025)Le, Bigo, Herremans \& Keller}]{le2025natural}
\bibinfo{author}{Le, D.-V.-T.}, \bibinfo{author}{Bigo, L.}, \bibinfo{author}{Herremans, D.}, \& \bibinfo{author}{Keller, M.} (\bibinfo{year}{2025}).
\newblock \bibinfo{title}{Natural language processing methods for symbolic music generation and information retrieval: a survey}.
\newblock {\it \bibinfo{journal}{ACM Computing Surveys}\/},  {\it \bibinfo{volume}{57}\/}, \bibinfo{pages}{1--40}.
\bibitem[{Liang et~al.(2020)Liang, Lei, Chan, Yang, Sun \& Chua}]{liang2020pirhdy}
\bibinfo{author}{Liang, H.}, \bibinfo{author}{Lei, W.}, \bibinfo{author}{Chan, P.~Y.}, \bibinfo{author}{Yang, Z.}, \bibinfo{author}{Sun, M.}, \& \bibinfo{author}{Chua, T.-S.} (\bibinfo{year}{2020}).
\newblock \bibinfo{title}{Pirhdy: Learning pitch-, rhythm-, and dynamics-aware embeddings for symbolic music}.
\newblock In {\it \bibinfo{booktitle}{Proceedings of the 28th ACM international conference on multimedia}\/} (pp. \bibinfo{pages}{574--582}).
\bibitem[{Lin et~al.(2024)Lin, Chen, Tang, Sha, Yang, Ju, Fan, Kang, Wu \& Meng}]{lin2024multi}
\bibinfo{author}{Lin, Z.}, \bibinfo{author}{Chen, J.}, \bibinfo{author}{Tang, B.}, \bibinfo{author}{Sha, B.}, \bibinfo{author}{Yang, J.}, \bibinfo{author}{Ju, Y.}, \bibinfo{author}{Fan, F.}, \bibinfo{author}{Kang, S.}, \bibinfo{author}{Wu, Z.}, \& \bibinfo{author}{Meng, H.} (\bibinfo{year}{2024}).
\newblock \bibinfo{title}{Multi-view midivae: Fusing track-and bar-view representations for long multi-track symbolic music generation}.
\newblock In {\it \bibinfo{booktitle}{ICASSP 2024-2024 IEEE International Conference on Acoustics, Speech and Signal Processing (ICASSP)}\/} (pp. \bibinfo{pages}{1--5}).
\newblock \bibinfo{organization}{IEEE}.
\bibitem[{Liu et~al.(2022)Liu, Dong, Cheng, Zhang, Li, Yu \& Sun}]{Liu2022symphonynet}
\bibinfo{author}{Liu, J.}, \bibinfo{author}{Dong, Y.}, \bibinfo{author}{Cheng, Z.}, \bibinfo{author}{Zhang, X.}, \bibinfo{author}{Li, X.}, \bibinfo{author}{Yu, F.}, \& \bibinfo{author}{Sun, M.} (\bibinfo{year}{2022}).
\newblock \bibinfo{title}{Symphony generation with permutation invariant language model}.
\newblock In {\it \bibinfo{booktitle}{Proceedings of the 23rd International Society for Music Information Retrieval Conference, {ISMIR} 2022, Bengaluru, India, December 4-8, 2022}\/} (pp. \bibinfo{pages}{551--558}).
\bibitem[{Lu et~al.(2023)Lu, Xu, Kang, Yu, Xing, Tan \& Bian}]{lu2023musecoco}
\bibinfo{author}{Lu, P.}, \bibinfo{author}{Xu, X.}, \bibinfo{author}{Kang, C.}, \bibinfo{author}{Yu, B.}, \bibinfo{author}{Xing, C.}, \bibinfo{author}{Tan, X.}, \& \bibinfo{author}{Bian, J.} (\bibinfo{year}{2023}).
\newblock \bibinfo{title}{Musecoco: Generating symbolic music from text}.
\newblock {\it \bibinfo{journal}{arXiv preprint arXiv:2306.00110}\/}, .
\bibitem[{Luo et~al.(2020)Luo, Yang, Ji \& Li}]{luo2020mg}
\bibinfo{author}{Luo, J.}, \bibinfo{author}{Yang, X.}, \bibinfo{author}{Ji, S.}, \& \bibinfo{author}{Li, J.} (\bibinfo{year}{2020}).
\newblock \bibinfo{title}{Mg-vae: Deep chinese folk songs generation with specific regional styles}.
\newblock In {\it \bibinfo{booktitle}{Proceedings of the 7th Conference on Sound and Music Technology (CSMT)}\/} (pp. \bibinfo{pages}{93--106}).
\bibitem[{Mukherjee \& Mulimani(2022)}]{mukherjee2022composeinstyle}
\bibinfo{author}{Mukherjee, S.}, \& \bibinfo{author}{Mulimani, M.} (\bibinfo{year}{2022}).
\newblock \bibinfo{title}{Composeinstyle: Music composition with and without style transfer}.
\newblock {\it \bibinfo{journal}{Expert Systems with Applications}\/},  {\it \bibinfo{volume}{191}\/}, \bibinfo{pages}{116195}.
\bibitem[{Payne(2019)}]{payne2019musenet}
\bibinfo{author}{Payne, C.} (\bibinfo{year}{2019}).
\newblock \bibinfo{title}{Musenet}.
\newblock \bibinfo{howpublished}{Retrieved from \url{https://openai. com/blog/musenet}. Accessed November 20, 2024}.
\bibitem[{Raffel(2016)}]{raffel2016learning}
\bibinfo{author}{Raffel, C.} (\bibinfo{year}{2016}).
\newblock {\it \bibinfo{title}{Learning-based methods for comparing sequences, with applications to audio-to-midi alignment and matching}\/}.
\newblock Ph.D. thesis Columbia University.
\bibitem[{Ren et~al.(2020)Ren, He, Tan, Qin, Zhao \& Liu}]{ren2020popmag}
\bibinfo{author}{Ren, Y.}, \bibinfo{author}{He, J.}, \bibinfo{author}{Tan, X.}, \bibinfo{author}{Qin, T.}, \bibinfo{author}{Zhao, Z.}, \& \bibinfo{author}{Liu, T.-Y.} (\bibinfo{year}{2020}).
\newblock \bibinfo{title}{Popmag: Pop music accompaniment generation}.
\newblock In {\it \bibinfo{booktitle}{Proceedings of the 28th ACM international conference on multimedia}\/} (pp. \bibinfo{pages}{1198--1206}).
\bibitem[{von R{\"{u}}tte et~al.(2023)von R{\"{u}}tte, Biggio, Kilcher \& Hofmann}]{von2023figaro}
\bibinfo{author}{von R{\"{u}}tte, D.}, \bibinfo{author}{Biggio, L.}, \bibinfo{author}{Kilcher, Y.}, \& \bibinfo{author}{Hofmann, T.} (\bibinfo{year}{2023}).
\newblock \bibinfo{title}{{FIGARO:} controllable music generation using learned and expert features}.
\newblock In {\it \bibinfo{booktitle}{The Eleventh International Conference on Learning Representations}\/} (pp. \bibinfo{pages}{1--18}).
\bibitem[{Takuya(1999)}]{takuya1999realtime}
\bibinfo{author}{Takuya, F.} (\bibinfo{year}{1999}).
\newblock \bibinfo{title}{Realtime chord recognition of musical sound: Asystem using common lisp music}.
\newblock In {\it \bibinfo{booktitle}{Proceedings of the International Computer Music Conference 1999, Beijing}\/}.
\bibitem[{Tan \& Herremans(2020)}]{tan2020music}
\bibinfo{author}{Tan, H.~H.}, \& \bibinfo{author}{Herremans, D.} (\bibinfo{year}{2020}).
\newblock \bibinfo{title}{Music fadernets: Controllable music generation based on high-level features via low-level feature modelling}.
\newblock In {\it \bibinfo{booktitle}{Proceedings of the 21th International Society for Music Information Retrieval Conference}\/} (pp. \bibinfo{pages}{109--116}).
\bibitem[{Tie et~al.(2024)Tie, Guo, Zhang, Tie, Qi \& Lu}]{tie2024hybrid}
\bibinfo{author}{Tie, Y.}, \bibinfo{author}{Guo, X.}, \bibinfo{author}{Zhang, D.}, \bibinfo{author}{Tie, J.}, \bibinfo{author}{Qi, L.}, \& \bibinfo{author}{Lu, Y.} (\bibinfo{year}{2024}).
\newblock \bibinfo{title}{Hybrid learning module-based transformer for multitrack music generation with music theory}.
\newblock {\it \bibinfo{journal}{IEEE Transactions on Computational Social Systems}\/},  (pp. \bibinfo{pages}{1--11}).
\bibitem[{Van Den~Oord et~al.(2017)Van Den~Oord, Vinyals et~al.}]{van2017neural}
\bibinfo{author}{Van Den~Oord, A.}, \bibinfo{author}{Vinyals, O.} et~al. (\bibinfo{year}{2017}).
\newblock \bibinfo{title}{Neural discrete representation learning}.
\newblock {\it \bibinfo{journal}{Advances in neural information processing systems}\/},  {\it \bibinfo{volume}{30}\/}.
\bibitem[{Vaswani et~al.(2017)Vaswani, Shazeer, Parmar, Uszkoreit, Jones, Gomez, Kaiser \& Polosukhin}]{vaswani2017attention}
\bibinfo{author}{Vaswani, A.}, \bibinfo{author}{Shazeer, N.}, \bibinfo{author}{Parmar, N.}, \bibinfo{author}{Uszkoreit, J.}, \bibinfo{author}{Jones, L.}, \bibinfo{author}{Gomez, A.~N.}, \bibinfo{author}{Kaiser, {\L}.}, \& \bibinfo{author}{Polosukhin, I.} (\bibinfo{year}{2017}).
\newblock \bibinfo{title}{Attention is all you need}.
\newblock {\it \bibinfo{journal}{Advances in neural information processing systems}\/},  {\it \bibinfo{volume}{30}\/}.
\bibitem[{Wang et~al.(2024{\natexlab{a}})Wang, Chen \& Li}]{wang2023continuous}
\bibinfo{author}{Wang, Y.}, \bibinfo{author}{Chen, M.}, \& \bibinfo{author}{Li, X.} (\bibinfo{year}{2024}{\natexlab{a}}).
\newblock \bibinfo{title}{Continuous emotion-based image-to-music generation}.
\newblock {\it \bibinfo{journal}{IEEE Transactions on Multimedia}\/},  {\it \bibinfo{volume}{26}\/}, \bibinfo{pages}{5670--5679}.
\bibitem[{Wang et~al.(2024{\natexlab{b}})Wang, Min \& Xia}]{wang2024whole}
\bibinfo{author}{Wang, Z.}, \bibinfo{author}{Min, L.}, \& \bibinfo{author}{Xia, G.} (\bibinfo{year}{2024}{\natexlab{b}}).
\newblock \bibinfo{title}{Whole-song hierarchical generation of symbolic music using cascaded diffusion models}.
\newblock In {\it \bibinfo{booktitle}{The Twelfth International Conference on Learning Representations, {ICLR} 2024, Vienna, Austria, May 7-11, 2024}\/} (pp. \bibinfo{pages}{1--21}).
\bibitem[{Wu \& Yang(2023)}]{wu2023musemorphose}
\bibinfo{author}{Wu, S.-L.}, \& \bibinfo{author}{Yang, Y.-H.} (\bibinfo{year}{2023}).
\newblock \bibinfo{title}{Musemorphose: Full-song and fine-grained piano music style transfer with one transformer vae}.
\newblock {\it \bibinfo{journal}{IEEE/ACM Transactions on Audio, Speech, and Language Processing}\/},  {\it \bibinfo{volume}{31}\/}, \bibinfo{pages}{1953--1967}.
\bibitem[{Yu et~al.(2022)Yu, Lu, Wang, Hu, Tan, Ye, Zhang, Qin \& Liu}]{yu2022museformer}
\bibinfo{author}{Yu, B.}, \bibinfo{author}{Lu, P.}, \bibinfo{author}{Wang, R.}, \bibinfo{author}{Hu, W.}, \bibinfo{author}{Tan, X.}, \bibinfo{author}{Ye, W.}, \bibinfo{author}{Zhang, S.}, \bibinfo{author}{Qin, T.}, \& \bibinfo{author}{Liu, T.-Y.} (\bibinfo{year}{2022}).
\newblock \bibinfo{title}{Museformer: Transformer with fine-and coarse-grained attention for music generation}.
\newblock {\it \bibinfo{journal}{Advances in Neural Information Processing Systems}\/},  {\it \bibinfo{volume}{35}\/}, \bibinfo{pages}{1376--1388}.
\bibitem[{Zeng et~al.(2021)Zeng, Tan, Wang, Ju, Qin \& Liu}]{zeng2021musicbert}
\bibinfo{author}{Zeng, M.}, \bibinfo{author}{Tan, X.}, \bibinfo{author}{Wang, R.}, \bibinfo{author}{Ju, Z.}, \bibinfo{author}{Qin, T.}, \& \bibinfo{author}{Liu, T.-Y.} (\bibinfo{year}{2021}).
\newblock \bibinfo{title}{Musicbert: Symbolic music understanding with large-scale pre-training}.
\newblock In {\it \bibinfo{booktitle}{Findings of the Association for Computational Linguistics: ACL-IJCNLP 2021}\/} (pp. \bibinfo{pages}{791--800}).
\bibitem[{Zhang et~al.(2022)Zhang, Zhang, Qiu, Wang \& Zhou}]{zhang2022structure}
\bibinfo{author}{Zhang, X.}, \bibinfo{author}{Zhang, J.}, \bibinfo{author}{Qiu, Y.}, \bibinfo{author}{Wang, L.}, \& \bibinfo{author}{Zhou, J.} (\bibinfo{year}{2022}).
\newblock \bibinfo{title}{Structure-enhanced pop music generation via harmony-aware learning}.
\newblock In {\it \bibinfo{booktitle}{Proceedings of the 30th ACM International Conference on Multimedia}\/} (pp. \bibinfo{pages}{1204--1213}).
\bibitem[{Zhang et~al.(2023)Zhang, Yu \& Takasu}]{zhang2023controllable}
\bibinfo{author}{Zhang, Z.}, \bibinfo{author}{Yu, Y.}, \& \bibinfo{author}{Takasu, A.} (\bibinfo{year}{2023}).
\newblock \bibinfo{title}{Controllable lyrics-to-melody generation}.
\newblock {\it \bibinfo{journal}{Neural Computing and Applications}\/},  {\it \bibinfo{volume}{35}\/}, \bibinfo{pages}{19805--19819}.
\bibitem[{Zhao et~al.(2024)Zhao, Xia, Wang \& Wang}]{zhao2023accomontage}
\bibinfo{author}{Zhao, J.}, \bibinfo{author}{Xia, G.}, \bibinfo{author}{Wang, Z.}, \& \bibinfo{author}{Wang, Y.} (\bibinfo{year}{2024}).
\newblock \bibinfo{title}{Structured multi-track accompaniment arrangement via style prior modelling}.
\newblock In {\it \bibinfo{booktitle}{The Thirty-eighth Annual Conference on Neural Information Processing Systems}\/} (pp. \bibinfo{pages}{1--32}).
\bibitem[{Zhou et~al.(2019)Zhou, Chu, Young \& Chen}]{Zhou2019BandNet}
\bibinfo{author}{Zhou, Y.}, \bibinfo{author}{Chu, W.}, \bibinfo{author}{Young, S.}, \& \bibinfo{author}{Chen, X.} (\bibinfo{year}{2019}).
\newblock \bibinfo{title}{Bandnet: {A} neural network-based, multi-instrument beatles-style {MIDI} music composition machine}.
\newblock In {\it \bibinfo{booktitle}{Proceedings of the 20th International Society for Music Information Retrieval Conference, {ISMIR} 2019, Delft, The Netherlands, November 4-8, 2019}\/} (pp. \bibinfo{pages}{655--662}).
\bibitem[{Zou et~al.(2022)Zou, Zou, Zhao, Zhang, Zhang \& Wang}]{zou2022melons}
\bibinfo{author}{Zou, Y.}, \bibinfo{author}{Zou, P.}, \bibinfo{author}{Zhao, Y.}, \bibinfo{author}{Zhang, K.}, \bibinfo{author}{Zhang, R.}, \& \bibinfo{author}{Wang, X.} (\bibinfo{year}{2022}).
\newblock \bibinfo{title}{Melons: generating melody with long-term structure using transformers and structure graph}.
\newblock In {\it \bibinfo{booktitle}{ICASSP 2022-2022 IEEE International Conference on Acoustics, Speech and Signal Processing (ICASSP)}\/} (pp. \bibinfo{pages}{191--195}).
\newblock \bibinfo{organization}{IEEE}.
\bibitem[{Zukowski \& Barthet(2023)}]{zukowski2023gtr}
\bibinfo{author}{Zukowski, Z.}, \& \bibinfo{author}{Barthet, M.} (\bibinfo{year}{2023}).
\newblock \bibinfo{title}{Gtr-ctrl: Instrument and genre conditioning for guitar-focused music generation with transformers}.
\newblock In {\it \bibinfo{booktitle}{Artificial Intelligence in Music, Sound, Art and Design: 12th International Conference}\/} (p. \bibinfo{pages}{260}).
\newblock volume \bibinfo{volume}{13988}.

\end{thebibliography}






\end{document}